\documentclass[11pt]{article}
\pdfoutput=1
\usepackage{jheppub}
\usepackage{amsmath,amssymb,amsfonts,graphicx,slashed,color,amsthm,mathtools,upgreek, enumerate, tensor, subfig}
\usepackage{arydshln}
\allowdisplaybreaks

\title{Modular Hamiltonians for Euclidean Path Integral States}
\author[a]{Srivatsan Balakrishnan}
\author[b]{\!, Onkar Parrikar}

\affiliation[a]{Department of Physics, University of Illinois, 1110 W. Green St., Urbana IL 61801, USA.} 
\affiliation[b]{Stanford Institute for Theoretical Physics, Department of Physics, Stanford University, CA 94305, USA.}

\emailAdd{sblkrsh2@illinois.edu}
\emailAdd{parrikar@stanford.edu}

\abstract{We study half-space/Rindler modular Hamiltonians for excited states created by turning on sources for local operators in the Euclidean path integral in relativistic quantum field theories. We derive a simple, manifestly Lorentzian formula for the modular Hamiltonian to all orders in perturbation theory in the sources. We apply this formula to the case of shape-deformed half spaces in the vacuum state, and obtain the corresponding modular Hamiltonian to all orders in the shape deformation in terms of products of half-sided null energy operators, i.e., stress tensor components integrated along the future and past Rindler horizons. In the special case where the shape deformation is purely null, our perturbation series can be re-summed, and agrees precisely with the known formula for vacuum modular Hamiltonians for null cuts of the Rindler horizon. Finally, we study some universal properties of modular flow (corresponding to Euclidean path integral states) of local operators inside correlation functions in conformal field theories. In particular, we show how the flow becomes the local boost in the limit where the operator being flowed approaches the entanglement cut.   

}

\date{September 2019}

\newcommand{\beq}{\begin{equation}}
\newcommand{\eeq}{\end{equation}}
\newcommand{\beqn}{\begin{eqnarray}}
\newcommand{\eeqn}{\end{eqnarray}}
\newcommand{\rni}{\text{I}}
\newcommand{\rnii}{\text{II}}

\begin{document}

\maketitle
\parskip=12pt
\section{Introduction}
The entanglement structure of states in quantum field theory has been a topic of active study in recent years. Of particular importance in this context is the notion of \emph{modular flow}. Given a state $\psi$ and a spatial subregion $R$, the modular Hamiltonian is defined as
\beq
K = -\ln\,\rho_{\psi,R}
\eeq
where $\rho_{\psi,R}$ is the reduced density matrix of $\psi$ over the subregion $R$: $\rho_{\psi,R} = \mathrm{Tr}_{R^c}|\psi\rangle\langle \psi|$. The modular Hamiltonian defines a natural flow on operators in the subregion by time-evolution, i.e., 
\beq
O \to e^{\frac{is}{2\pi}K}O e^{-\frac{is}{2\pi}K},
\eeq
called modular flow. While at face value the modular Hamiltonian and flow may seem rather abstract, they have several interesting properties and applications. For instance, the modular Hamiltonian has a monotonicity property under inclusions, which was used in \cite{Faulkner:2016mzt} to prove the averaged null energy condition in relativistic quantum field theories. Correlation functions of modular-flowed operators have rather general analyticity properties, which were leveraged in \cite{Balakrishnan:2017bjg, Ceyhan:2018zfg} to prove a stronger constraint called the quantum null energy condition. Further, modular flow plays a crucial role in subregion duality in the AdS/CFT correspondence. For instance, it was argued in \cite{Jafferis:2015del} that the quantum-generalized Ryu-Takayanagi (RT) formula \cite{Ryu:2006bv, Hubeny:2007xt, Faulkner:2013ana, Engelhardt:2014gca} implies that the modular Hamiltonian for a boundary subregion (projected to the ``code subspace'') is equal to the modular Hamiltonian of bulk quantum fields in the corresponding entanglement wedge, up to a term localized to the Ryu-Takayanagi surface. This equivalence of bulk and boundary modular flows is a rather powerful statement about the matching of bulk and boundary ``algebraic symmetries'', and has led to general arguments \cite{Dong:2016eik, Harlow:2016vwg} as well as concrete prescriptions \cite{Faulkner:2017vdd, Faulkner:2018faa, Cotler:2017erl, Czech:2019vih} for bulk-reconstruction within the entanglement wedge of a boundary subregion using modular flow.\footnote{These can be viewed as a natural generalization of the HKLL construction \cite{2006PhRvD..74f6009H}, which works inside the (generally smaller) causal wedge.} Properties of modular flow have also been used in \cite{Nozaki:2013vta, Faulkner:2013ica, Faulkner:2017tkh, Haehl:2017sot, Lewkowycz:2018sgn} to argue that any bulk geometry holographically dual to a CFT state satisfying the RT formula must necessarily satisfy the non-linear Einstein equation.  

Despite this, useful expressions for modular Hamiltonians are rather hard to come by in general spacetime dimension. The notable exceptions to this are special subregions such as the half-space/Rindler wedge \cite{Bisognano:1976za} or null deformations thereof \cite{Faulkner:2016mzt, Wall:2011hj, Bousso:2014uxa, Casini:2017roe, Balakrishnan:2017bjg, Lashkari:2017rcl, Koeller:2017njr} in the vacuum state of relativistic quantum field theories, and ball shaped regions in the vacuum state of conformal field theories \cite{Casini:2011kv, Wong:2013gua} (see also \cite{Cardy:2016fqc} for results on modular Hamiltonians in 2d CFTs). But for more general states and subregions, not many accessible tools exist to study modular flow, especially outside the purview of free field theories \cite{Casini:2009sr, Arias:2018tmw, Wong:2018svs} and AdS/CFT (although, see \cite{Lashkari:2015dia} for a replica trick approach). In this paper, we seek to alleviate this situation by giving a more or less simple and \emph{manifestly Lorentzian} formula for the half-space/Rindler modular Hamiltonian for a particular class of excited states in relativistic quantum field theories. The states in question are created by turning on a source (with a small amplitude) for some local operator $O$ in the Euclidean path integral. These Euclidean path integral states are of direct relevance in AdS/CFT since they have geometric bulk duals \cite{Skenderis:2008dg, Botta-Cantcheff:2015sav, Marolf:2017kvq}, and thus constitute an important class of excited states for which we would like to construct the modular Hamiltonian. Our approach to this will be to treat the source perturbatively (building on previous work by \cite{Faulkner:2016mzt, Faulkner:2017tkh, Rosenhaus:2014woa, Rosenhaus:2014zza, Faulkner:2014jva, Lewkowycz:2014jia, Faulkner:2015csl, Speranza:2016jwt, Sarosi:2016atx, Sarosi:2017rsq, Lashkari:2018oke, Lashkari:2018tjh, Ugajin:2018rwd, Haehl:2019fjz}). We claim that to all orders in this perturbation theory, the modular Hamiltonian is given by:
\beq \label{ModHamFinalIntro}
K_{\lambda}= c_{\lambda}+K +\sum_{n=1}^{\infty}\frac{1}{n!}\delta^nK ,
\eeq
\beq
\delta^nK=n!\frac{(-i)^{n-1}}{(2\pi)^{n-1}}\int d\mu_n \int_{-\infty}^{\infty}ds_1\cdots\int_{-\infty}^{\infty}ds_n\, f_{(n)}(s_1+i\tau_1\cdots,s_n+i\tau_n)O(s_1,Y_1)\cdots O(s_n, Y_n),
\eeq
where $K$ is the vacuum modular Hamiltonian, $\tau$ is the angular coordinate around the entanglement cut which parametrizes vacuum modular flow, and $Y$ denotes spatial coordinates on the half-space subregion. Further,  $d\mu_n$ is defined as
\beq
d\mu_n = \prod_{i=1}^n d\tau_i d^{d-1}Y_i\,\lambda(\tau_i,Y_i)
\eeq
and contains $n$ powers of the source $\lambda$, and $c_{\lambda}$ is a constant. The operators appearing in this expansion are defined as
\beq
O(s,Y) = e^{\frac{is}{2\pi}K}O(0,Y) e^{-\frac{is}{2\pi}K},
\eeq
where once again $K$ is the vacuum modular Hamiltonian. Finally, we have defined the functions
\beq
f_{(n)}(s_1,\cdots s_n) = \frac{1}{2^{n+1}}\frac{1}{\sinh(\frac{s_1}{2})\sinh(\frac{s_2-s_1}{2})\cdots\sinh(\frac{s_n-s_{n-1}}{2})\sinh(\frac{s_n}{2})}.
\eeq
We emphasize that our result \eqref{ModHamFinalIntro} for the modular Hamiltonian is manifestly Lorentzian, i.e., the operators appearing in it do not involve Euclidean/imaginary modular flow, but are only flowed in real modular time. In fact, this feature is \emph{not} a priori obvious during the intermediate stages of the calculation (where Euclidean modular time does appear) and in the end hinges on a rather surprising cancellation between various terms! This property is crucial, however, especially in using the final answer for computing entanglement or relative entropies for these excited states. Had our expression involved Euclidean modular flow, the resulting entanglement/relative entropies would contain spurious out-of-Euclidean-time orderings inside correlation functions, which are singular -- the fact that our formula is entirely written in terms of Lorentzian operators avoids these problems (see discussion around equation \eqref{entropy} for details). Additionally, in the context of AdS/CFT, we view this feature as a first step towards a fully Lorentzian understanding of the Ryu-Takayanagi formula. We will derive equation \eqref{ModHamFinalIntro} in section \ref{sec:sec2}. Similar perturbative expansions for the modular Hamiltonian have appeared before in \cite{Sarosi:2017rsq, Lashkari:2018oke, Lashkari:2018tjh}, but the key difference here is that our formula is completely Lorentzian.\footnote{There are other differences as well: firstly, the excited states considered in \cite{Sarosi:2017rsq, Lashkari:2018oke} are not Euclidean path integral states. Secondly, while the analogs of the $s$-integrals in the formulas appearing in these papers are Lorentzian, the modular Hamiltonian is not, i.e., it is not clear that the $s$-integrals allow one to get rid of the Euclidean modular evolution in $\rho$. This leads to operator-ordering issues in \cite{Sarosi:2017rsq}, which can sometimes be circumvented by intricate contour deformations. On the other hand, the authors of \cite{Lashkari:2018oke, Lashkari:2018tjh} express the perturbation series in terms of a bounded operator they denote $\boldsymbol{\delta}$. However, for Euclidean path integral states, $\boldsymbol{\delta}$ itself has an expansion in terms of local operators, making this formalism tedious to apply.} 

To demonstrate the utility of these results, we apply \eqref{ModHamFinalIntro} to the case of shape-deformed half-spaces in the vacuum state (which can be treated as turning on a particular source for the stress tensor in the Euclidean path integral) in section \ref{sec:sec3}, and give an expression (see equation \eqref{ModHamShape}) for the modular Hamiltonian to all orders in the shape deformation of the entanglement cut. This involves various products of half-sided null energy operators, namely stress tensor components $T_{++}$, $T_{+-}$ and $T_{--}$ integrated along the future and past null horizons of the subregion, and is essentially an all-orders generalization of the first order result derived in \cite{Faulkner:2016mzt}.\footnote{A similar result for the entanglement entropy of shape-deformed half-spaces was derived independently by Tom Faulkner in unpublished work.}  In the special case where the shape deformation is purely null, our perturbative expansion simplifies and the series can be re-summed. The result precisely agrees with the known formula for the vacuum modular Hamiltonian for null cuts of the Rindler horizon \cite{Wall:2011hj, Bousso:2014uxa, Faulkner:2016mzt, Casini:2017roe, Lashkari:2017rcl, Balakrishnan:2017bjg, Koeller:2017njr}. Finally, in section \ref{sec:sec4}, we apply our formula (at leading orders) to demonstrate how modular flow corresponding to Euclidean path-integral states becomes a local boost (i.e., identical to the vacuum modular flow) in the limit where the operator being flowed approaches the entanglement cut, a property which has generally been expected to be true \cite{Faulkner:2017tkh, Lewkowycz:2018sgn}.  

\section{Modular perturbation theory for Euclidean Path-integral States}\label{sec:sec2}
In this section, we will derive the half-space/Rindler modular Hamiltonian for Euclidean path integral states, building on previous work by \cite{Faulkner:2016mzt, Faulkner:2017tkh, Rosenhaus:2014woa, Rosenhaus:2014zza, Faulkner:2014jva, Lewkowycz:2014jia, Faulkner:2015csl, Speranza:2016jwt, Sarosi:2016atx, Sarosi:2017rsq, Lashkari:2018oke, Lashkari:2018tjh, Ugajin:2018rwd, Haehl:2019fjz}. These states are constructed as follows: the vacuum state in a general quantum field theory can be constructed in terms of the Euclidean path integral for Euclidean times $x^0_E <0$:
\beq
\langle \varphi(\mathbf{x}) | 0\rangle = \int^{\phi(0^-,\mathbf{x}) = \varphi(\mathbf{x})} D\phi\,e^{-\int_{-\infty}^0 dx^0_E\int d^{d-1}\mathbf{x}\,\mathcal{L}[\phi]},
\eeq
where we have collectively denoted the elementary fields in the path integral as $\phi$, $\mathbf{x}\in \mathbb{R}^{d-1}$ is a coordinate on the $x_E^0 = 0$ slice, and the boundary condition $\varphi(\mathbf{x})$ at $x^0_E =0$ provides a basis for the Hilbert space. Now we can consider a natural class of excited states by turning on sources in this path integral for local operators:
\beq
\langle \varphi(\mathbf{x}) | \psi[\lambda_{(0)}^{\alpha}]\rangle = \int^{\phi(0^-,\mathbf{x}) = \varphi(\mathbf{x})} D\phi\,e^{-\int_{-\infty}^0 dx^0_E\int d^{d-1}\mathbf{x}\,\left(\mathcal{L}[\phi]-\sum_\alpha\lambda^{\alpha}_{(0)}(x)O_{\alpha}(x)\right)},
\eeq
where we have denoted $x=(x^0_E,\mathbf{x})$. These are the class of excited states we will be interested in. For simplicity, we will focus on the case where only one \emph{Hermitian} operator $O$ is turned on, but generalizing to the case of multiple operators is straightforward. We will additionally take our subregion to be the half space $R=\left\{\mathbf{x}=(x^1,\vec{x}) | x^1 \geq 0\right\}$, with $\vec{x}$ denoting the transverse directions along the entanglement cut. This has the advantage that the vacuum modular Hamiltonian is local \cite{Bisognano:1976za} and given by the boost operator restricted to $R$:
\beq\label{vacModHam}
K = 2\pi\int d^{d-2}\vec{x}\int_0^{\infty}dx^1\,x^1T_{00}(0,x^1,\vec{x}) +\text{const}. 
\eeq
In conformal field theories, our arguments will also work for ball-shaped regions, as these can be conformally mapped to a half-space \cite{Casini:2011kv}. 

\subsection{Reduced density matrix}
\begin{figure}
    \centering
    \includegraphics[height=5cm]{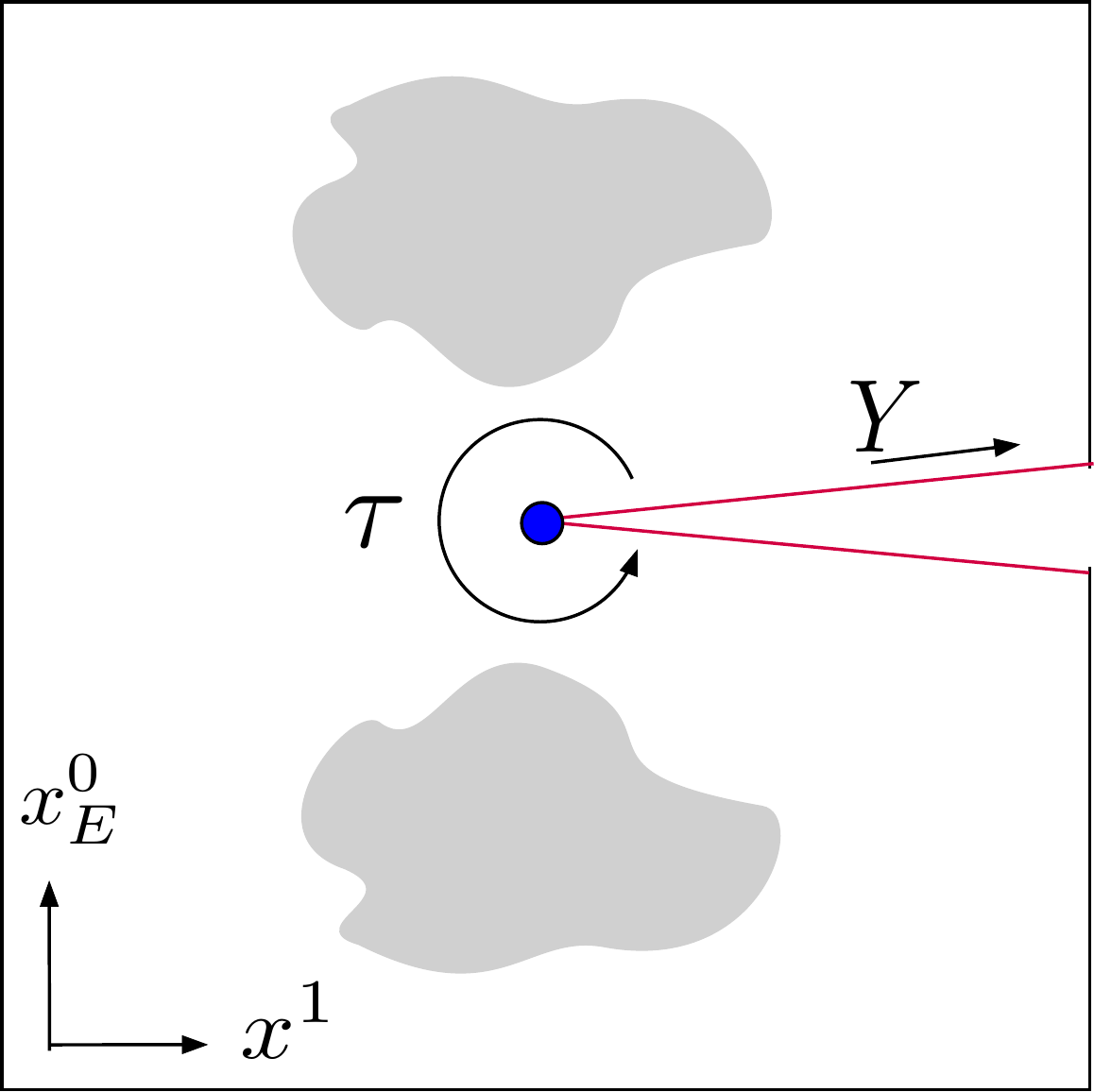}
    \caption{\small{The reduced density matrix is obtained by performing the Euclidean path integral over $\mathbb{R}^d$ with a cut along the subregion $R$ (shown in red). The shaded blobs denote regions where sources are turned on; note that they are time-reflection symmetric. Transverse directions are suppressed. }}
    \label{fig:rdm}
\end{figure}

Our next task is to construct the reduced density matrix over $R$ for this class of excited states. Following standard arguments, this is given by the following Euclidean path-integral over $\mathbb{R}^d$ with a cut along $R$:
\beq
\langle \varphi_-(Y) | \widehat\rho_{\lambda} | \varphi_+(Y)\rangle = \frac{1}{Z_{\lambda}}\int^{\phi(0^-,Y) = \varphi_-(Y)}_{\phi(0^+,Y) = \varphi_+(Y)} D\phi\,e^{-\int d^{d}x\,\left(\mathcal{L}[\phi]-\lambda(x)O(x)\right)},
\eeq
where we have introduced the coordinates $x=(\tau, Y)$ with $\tau$ being the angular coordinate in the $(x^0_E,x^1)$ plane and $Y=(x^1>0,\vec{x})$ are the remaining coordinates on the subregion $R$ (see figure \ref{fig:rdm}). The new source $\lambda$ is given in terms of $\lambda_{(0)}$ as follows:
\beq\label{sources}
\lambda(x^0_E, \mathbf{x}) = \begin{cases} \lambda_{(0)}(x^0_E, \mathbf{x}) & \cdots \;\;\; x_E^0 <0 \\ \lambda^*_{(0)}(-x^0_E, \mathbf{x}) & \cdots \;\;\; x_E^0 >0. \end{cases}
\eeq
Finally, the normalization $Z_{\lambda}$ is the Euclidean partition function on $\mathbb{R}^d$ in the presence of the source $\lambda$, and ensures that $\mathrm{Tr}\,\widehat{\rho}_{\lambda}=1$. 

Now we can expand this density matrix in powers of $\lambda$. Since we are eventually interested in taking the log of the reduced density matrix, it suffices to consider the unnormalized density matrix $\rho_{\lambda}$, with $\widehat\rho_{\lambda}= \frac{1}{Z_{\lambda}} \rho_{\lambda}$; the normalization simply adds an overall (source-dependent) constant piece to the modular Hamiltonian. It is easy to verify that this expansion takes the form
\beq\label{RDMexp}
\rho_{\lambda}=\rho + \sum_{n=1}^{\infty}\frac{1}{n!}\delta^n\rho,\;\;\delta^n\rho=\int d\mu_n\,\rho\mathcal{T}\left[O(\tau_1,Y_1)\cdots O(\tau_n,Y_n)\right]
\eeq
where 
\beq\label{measure}
d\mu_n = \prod_{i=1}^n d\tau_id^{d-1}Y_i\,\lambda(\tau_i,Y_i),
\eeq
and $\mathcal{T}$ stands for time-ordering in the Euclidean time coordinate $\tau$. Note that $\tau$ can be thought of as Euclidean modular time with respect to the vacuum modular Hamiltonian. This allows us to view the operators appearing above as Heisenberg operators with respect to Euclidean (vacuum) modular flow:
\beq\label{Heisenberg}
O(\tau, Y) = e^{\frac{\tau}{2\pi} K}O(0,Y) e^{-\frac{\tau}{2\pi} K},
\eeq
where $K$ is the vacuum modular Hamiltonian for the half-space $R$ defined in equation \eqref{vacModHam}. Although the expansion \eqref{RDMexp} is true more generally with a suitable choice of the angular coordinate $\tau$, \eqref{Heisenberg} is only true in cases where vacuum modular flow is local.\footnote{In the case of ball-shaped regions in conformal field theories, one must append suitable conformal factors to \eqref{Heisenberg}.} 

\subsection{Modular Hamiltonian}
Next, we want to construct the perturbative expansion of $K_{\lambda}=-\ln\,\widehat{\rho}_{\lambda}$. We claim that the final result is 
\beq \label{ModHamFinal}
K_{\lambda}= c_{\lambda}+K +\sum_{n=1}^{\infty}\frac{1}{n!}\delta^nK ,
\eeq
\beq
\delta^nK=n!\frac{(-i)^{n-1}}{(2\pi)^{n-1}}\int d\mu_n \int_{-\infty}^{\infty}ds_1\cdots\int_{-\infty}^{\infty}ds_n\, f_{(n)}(s_1+i\tau_1\cdots,s_n+i\tau_n)O(s_1,Y_1)\cdots O(s_n, Y_n),
\eeq
where $d\mu_n$ was defined in \eqref{measure} and contains $n$ powers of the source $\lambda$, and $c_{\lambda}=\ln\,\frac{Z_{\lambda}}{Z_0}$ is a constant. The operators appearing in this expansion are defined as
\beq
O(s,Y) = e^{\frac{is}{2\pi}K}O(0,Y) e^{-\frac{is}{2\pi}K},
\eeq
where once again $K$ is the vacuum modular Hamiltonian, and are purely Lorentzian. Finally, we have defined the functions
\beq
f_{(n)}(s_1,\cdots s_n) = \frac{1}{2^{n+1}}\frac{1}{\sinh(\frac{s_1}{2})\sinh(\frac{s_2-s_1}{2})\cdots\sinh(\frac{s_n-s_{n-1}}{2})\sinh(\frac{s_n}{2})}.
\eeq
It is easy to check that the functions $f_{(n)}$ have precisely the right symmetry properties (together with equation \eqref{sources}) to make each term in this expansion Hermitian. As another check, it can be verified that if the source $\lambda$ is $\tau$-independent (i.e., if the source preserves the boost symmetry around the entanglement cut), then one can perform the $\tau$ integrals in \eqref{ModHamFinal}. The only term which survives after these integrations is the $n=1$ term, where the $\tau$ integral leads to a contact term proportional to $\delta(s_1)$; this precisely combines with the leading vacuum modular Hamiltonian to give the new boost operator in the presence of $\lambda$, thus leading to a local modular Hamiltonian, as expected.  

We would like to emphasize the fact that equation \eqref{ModHamFinal} is manifestly Lorentzian -- this property is crucial, especially for computing entanglement or relative entropies for these excited states. For example, the entanglement entropy is given by
\beq\label{entropy}
S_{EE}= \sum_{n=0}^{\infty}\sum_{m=0}^n\frac{1}{m!(n-m)!}\mathrm{Tr}\,\delta^m\widehat{\rho}_{\lambda}\delta^{n-m}K_{\lambda}.
\eeq
From equation \eqref{RDMexp}, $\delta^m\widehat{\rho}_\lambda$ involves a (Euclidean) time-ordered product of operators, while $\delta^{n-m}K_{\lambda}$ is Lorentzian. Thus, the trace in each of these terms is Euclidean time-ordered, and can be written as a Euclidean correlation function. Had our expression involved Euclidean modular flow, the resulting entanglement/relative entropies would contain spurious out-of-Euclidean-time orderings inside the trace, which are singular -- the fact that our formula is entirely written in terms of Lorentzian operators avoids these problems. 

The rest of this section will be devoted to proving equation \eqref{ModHamFinal}. Our starting point will be the general formula\footnote{As mentioned previously, we will compute $-\log$ of the un-normalized density matrix. Below, we will abuse notation slightly and simply call this object $K_{\lambda}$. The true modular Hamiltonian is $-\log$ of the normalized density matrix. The normalization only contributes an additive constant which can be added to the final result at the end.}
\beq \label{exp1}
\delta K_{\lambda} = \int_{-\infty+i\theta}^{\infty+i\theta}\frac{ds}{4\sinh^2(s/2)}e^{isK_{\lambda}/2\pi}\rho_{\lambda}^{-1}\delta \rho_{\lambda} e^{-isK_{\lambda}/2\pi},\;\;\cdots\;\; (0<\theta < 2\pi)
\eeq
where $\delta$ stands for a derivative with respect to $\lambda$. This equation is true of any general family of density matrices parametrized by $\lambda$, and can be derived as follows: 
\beqn
\epsilon\delta K_{\lambda} &= & -\ln(\rho_{\lambda} + \epsilon\delta \rho_{\lambda}) + \ln \rho_{\lambda}  + O(\epsilon^2)\nonumber\\
&= & -\ln[\rho_{\lambda}(1 + \epsilon \rho_{\lambda}^{-1}\delta\rho_{\lambda})] + \ln \rho_{\lambda}  + O(\epsilon^2)\nonumber\\
&= & -\ln[e^{-K_{\lambda}}e^{ \rho_{\lambda}^{-1}\delta\rho_{\lambda}}] + \ln \rho_{\lambda}  + O(\epsilon^2).
\eeqn
Using the Baker-Campbell-Hausdorff formula in the first term above gives
\beq
\delta K_{\lambda} = -\sum_{n=0}^{\infty}(-1)^n\frac{B_n}{n!}\left[K_{\lambda}\left[K_{\lambda},\cdots\left[K_{\lambda},\rho_{\lambda}^{-1}\delta\rho_{\lambda}\right]\cdots\right]\right],
\eeq
where there are $n$ commutators with $K_{\lambda}$ on the right hand side above and $B_n = -n\zeta(1-n)$ are the Bernoulli numbers. Using the following integral representation (which can be proved using Cauchy's residue theorem):
\beq
B_n = -\frac{(-i)^n}{(2\pi)^n}\int_{-\infty+i\theta}^{\infty+i\theta}\frac{ds}{4\sinh^2(s/2)}s^n,\;\;\;\; \cdots \;\;\;(0 < \theta < 2\pi) 
\eeq
then gives the desired result in equation \eqref{exp1}. Equation \eqref{exp1} is completely general; we have not assumed anything (such as, for instance, locality) about $K_{\lambda}$. We have, however, assumed that $\rho_{\lambda}^{-1}\delta \rho_{\lambda}$ is well-defined, which we expect to be true in the case of Euclidean path-integral states. Also note that the parameter $\theta$ here is arbitrary, as long as it lies within the specified range. This freedom is going to be crucial in what follows. 

Our strategy to obtain the $n$th term in the expansion for the modular Hamiltonian will be to take $(n-1)$ more derivatives of equation \eqref{exp1} with respect to $\lambda$ and then set $\lambda=0$. Let us first see how this works at lower orders in perturbation theory, and we will then give an inductive proof for general $n$.

\subsection*{First order}
At first order, we can evaluate \eqref{exp1} at $\lambda=0$, which gives 
\beq
\delta K_{\lambda}\Big|_{\lambda=0} = \int_{-\infty+i\theta}^{\infty+i\theta}\frac{ds}{4\sinh^2(s/2)}e^{isK/2\pi}\rho^{-1}\delta \rho e^{-isK/2\pi},\;\;\cdots\;\; (0<\theta < 2\pi)
\eeq
From equation \eqref{RDMexp}, we have
\beq
\rho^{-1}\delta \rho = \int d\tau_1 d^{d-1}Y_1\,\lambda(\tau_1,Y_1)O(\tau_1, Y_1).
\eeq
Now by choosing the parameter $\theta = \tau_1$, we can completely remove all the Euclidean modular time dependence from within the operator. This gives a manifestly Lorentzian expression for the modular Hamiltonian at this order:
\beq \label{modHam1}
\delta K= \int d\tau_1 d^{d-1}Y_1\lambda(\tau_1,Y_1) \int_{-\infty}^{\infty}ds_1\, f_{(1)}(s_1+i\tau_1)O(s_1,Y_1),
\eeq
\beq
f_{(1)}(s) = \frac{1}{4\sinh^2(s/2)}.
\eeq
\subsection*{Second order}
Now we take one more derivative of equation \eqref{exp1}:
\beqn\label{exp2}
\delta^2K &=&  \int_{-\infty+i\theta}^{\infty+i\theta}\frac{ds}{4\sinh^2(s/2)}\Big\{\delta\left[e^{isK/2\pi}\right]\rho^{-1}\delta \rho e^{-isK/2\pi}+e^{isK/2\pi}\rho^{-1}\delta \rho \delta\left[e^{-isK/2\pi}\right]\nonumber\\
&+&e^{isK/2\pi}\delta\left[\rho^{-1}\delta \rho\right] e^{-isK/2\pi}\Big\}.
\eeqn
We will refer to the three terms appearing above as $T_1(0)$, $T_2(0)$ and $T(1)$ respectively, where the number in the parentheses denotes how many derivatives act on $\rho^{-1}\delta \rho$. Let us first consider the $T_1(0)$ term. We will begin by setting the contour parameter $\theta = \epsilon > 0$, i.e., infinitesimally small but positive, in this computation. We need to work out $\delta\left[e^{isK/2\pi}\right]$. Since $s$ is almost real, then we get
\beq \label{expder}
\delta\left[e^{isK/2\pi}\right]= \frac{i}{2\pi}\int_0^s dt\,e^{itK/2\pi}\delta K e^{-itK/2\pi}e^{isK/2\pi}
\eeq
up to corrections of $O(\epsilon)$. Using our previously derived formula \eqref{modHam1} for $\delta K$, we find
\beqn
\delta\,e^{\frac{isK}{2\pi}}&=&\frac{i}{2\pi}\int d\tau_1d^{d-1}Y_1\lambda(\tau_1,Y_1)\int_0^{s}dt\int_{-\infty+i\tau_1}^{\infty+i\tau_1}\frac{ds_1}{4\sinh^2(\frac{s_1}{2})}  e^{\frac{i(t+s_1)K}{2\pi}}O_1 e^{-\frac{i(t+s_1)K}{2\pi}}e^{\frac{isK}{2\pi}}\nonumber\\
&=&\frac{i}{2\pi}\int d\tau_1d^{d-1}Y_1\lambda(\tau_1,Y_1)\int_0^{s}dt\int_{-\infty+i\tau_1}^{\infty+i\tau_1}\frac{ds_1}{4\sinh^2(\frac{s_1-t}{2})}  e^{\frac{is_1K}{2\pi}}O_1 e^{-\frac{is_1K}{2\pi}}e^{\frac{isK}{2\pi}},
\eeqn
where we have used the convenient notation 
$$O_i = O(\tau_i, Y_i).$$ 
Next, we exchange the order of $t$ and $s_1$ integrations. This allows us to perform the $t$ integration:
\beq
\int_0^{s} dt\,\frac{1}{4\sinh^2(\frac{s_1-t}{2})} = \frac{1}{2}\frac{\sinh(s/2)}{\sinh(s_1/2)\sinh((s_1-s)/2)}.
\eeq
We therefore conclude that 
\beq
\delta\,e^{\frac{isK}{2\pi}}=\frac{-i}{2\pi}\int d\tau_1d^{d-1}Y_1\lambda(\tau_1,Y_1)\int_{-\infty+i\tau_1}^{\infty+i\tau_1}ds_1 \frac{\sinh(\frac{s}{2})}{2\sinh(\frac{s_1}{2})\sinh(\frac{s-s_1}{2})}  e^{\frac{is_1K}{2\pi}}O_1 e^{-\frac{is_1K}{2\pi}}e^{\frac{isK}{2\pi}},
\eeq
and so the first term in \eqref{exp2} is given by (relabeling $s \to s_2$)
\beq
T_1(0)=\frac{-i}{2\pi}\int_{-\infty+i\epsilon}^{\infty+i\epsilon}ds_2\int_{-\infty+i\tau_1}^{\infty+i\tau_1} ds_1\,f_{(2)}(s_1,s_2)  e^{\frac{is_1K}{2\pi}}O_1 e^{-\frac{is_1K}{2\pi}}e^{\frac{is_2K}{2\pi}}O_2 e^{-\frac{is_2K}{2\pi}},
\eeq
where
\beq
d\mu_2 = \int d\tau_1 dY_1 \lambda(\tau_1,Y_1) \int d\tau_2 dY_2 \lambda(\tau_2,Y_2),
\eeq
and
\beq
f_{(2)}(s_1,s_2) = \frac{1}{8\sinh(s_1/2)\sinh(\frac{s_2-s_1}{2})\sinh(s_2/2)}.
\eeq

Note that since the $s_1$ contour has $\text{Im}\,s_1 = \tau_1$, then the corresponding operator is purely Lorentzian, i.e., has no flow in Euclidean time. For $O_2$ as well, we can deform the $s_2$ contour of integration to place it at $\text{Im}\,s_2 = \tau_2$ in order to get rid of the Euclidean time dependence in it. This deformation is innocuous if $\tau_2 < \tau_1$. But if $\tau_2 > \tau_1$, then we cross a pole in $f_{(2)}(s_1,s_2)$ and pick up an additional contribution.\footnote{Here we have exchanged the order of $s_1$ and $s_2$ integrals.} So all in all, we get
\beqn
T_1(0) &=&\frac{-i}{2\pi}\int d\mu_2 \int_{-\infty}^{\infty}ds_1\int_{-\infty}^{\infty} ds_2\,f_{(2)}(s_1+i\tau_1,s_2+i\tau_2)  O(s_1,Y_1) O(s_2,Y_2)\nonumber\\
&+&\int d\mu_2\,\Theta(\tau_2-\tau_1)\int_{-\infty+i\tau_1}^{\infty+i\tau_1}\frac{ds}{4\sinh^2(s/2)}e^{isK/2\pi}O(\tau_1,Y_1)O(\tau_2,Y_2)e^{-isK/2\pi}.
\eeqn

We can similarly work out the $T_2(0)$ term, and we get
\beqn
T_2(0)&=&\frac{-i}{2\pi}\int d\mu_2 \int_{-\infty}^{\infty}ds_1\int_{-\infty}^{\infty} ds_2\,f_{(2)}(s_1+i\tau_1,s_2+i\tau_2)  O(s_1,Y_1) O(s_2,Y_2)\nonumber\\
&-&\int d\mu\,\Theta(\tau_2-\tau_1)\int_{-\infty+i\tau_1}^{\infty+i\tau_1}\frac{ds}{4\sinh^2(s/2)}e^{isK/2\pi}O(\tau_2,Y_2)O(\tau_1,Y_1)e^{-isK/2\pi}.
\eeqn
On the other hand, the $T(1)$ term is easy to calculate using \eqref{RDMexp}:
\beq
T(1) = \int d\mu\,\int_{-\infty+i\epsilon}^{\infty+i\epsilon}\frac{ds}{4\sinh^2(s/2)}e^{isK/2\pi}\left\{\mathcal{T}[O(\tau_1)O(\tau_2)]-O(\tau_1,Y_1)O(\tau_2,Y_2)\right\}e^{-isK/2\pi}.
\eeq
Now through a simple contour deformation, we see that the second line of $T_1(0)$ together with the second line of $T_2(0)$ cancels with $T(1)$! This cancellation crucially removes all Euclidean modular flow from our expressions, and we are left with a manifestly Lorentzian result:
\beq \label{secondorder}
\delta^2K = 2!\frac{-i}{2\pi}\int d\mu \int_{-\infty}^{\infty}ds_1\int_{-\infty}^{\infty} ds_2\,f_{(2)}(s_1+i\tau_1,s_2+i\tau_2)  O(s_1,Y_1) O(s_2,Y_2).
\eeq
This is our final expression for the second derivative of the modular Hamiltonian, in agreement with our claim \eqref{ModHamFinal}.

One can proceed with this calculation systematically at higher orders. At the third order, for instance, we would first need to calculate $\delta^2 e^{\frac{isK}{2\pi}}$ (and its complex conjugate), and then use it together with lower order results to compute $\delta^3 K$. Once again, we find a cancellation between all terms involving Euclidean modular flow and the final answer agrees with \eqref{ModHamFinal}. More generally, at the $n$th order one needs to calculate $\delta^{n-1} e^{\frac{isK}{2\pi}}$ and its conjugate using lower order results, and then use these to evaluate $\delta^n K$. In doing this, one observes the following pattern:
\beqn \label{induction1}
\delta^{n-1}e^{\frac{isK}{2\pi}}&=& \frac{(n-1)!(-i)^{n-1}}{(2\pi)^{n-1}}\int d\mu_{n-1}\int_{i\tau_1} ds_1\cdots \int_{i\tau_{n-1}} ds_{n-1}\frac{f_{(n)}(s_1,\cdots, s_{n-1},s)}{f_{(1)}(s)}\\
&\times & e^{\frac{is_1K}{2\pi}}O(\tau_1,Y_1)e^{-\frac{is_1K}{2\pi}}e^{\frac{is_2K}{2\pi}}O(\tau_2,Y_2)e^{-\frac{is_2K}{2\pi}}\cdots e^{\frac{is_{n-1}K}{2\pi}}O(\tau_{n-1},Y_{n-1})e^{-\frac{is_{n-1}K}{2\pi}}e^{\frac{isK}{2\pi}},\nonumber
\eeqn
and
\beqn \label{induction2}
\delta^{n}K&=&\frac{n!(-i)^{n-1}}{(2\pi)^{n-1}}\int d\mu_n\int_{i\tau_1} ds_1\cdots \int_{i\tau_n}ds_{n} f_{(n)}(s_1,\cdots,s_n)\nonumber\\
&\times &e^{\frac{is_1K}{2\pi}}O(\tau_1,Y_1)e^{-\frac{is_1K}{2\pi}}e^{\frac{is_2K}{2\pi}}O(\tau_2,Y_2)e^{-\frac{is_2K}{2\pi}}\cdots e^{\frac{is_{n}K}{2\pi}}O(\tau_{n},Y_{n})e^{-\frac{is_{n}K}{2\pi}}.
\eeqn
So with this pattern in hand, we can now give a general proof of these formulas by the method of induction.

\subsection{Proof at general order by induction}
We have already proved equations \eqref{induction1} and \eqref{induction2} for $n=2$. In order to proceed by the method of induction, let's assume we have proved these formulas up to some positive integer $n-1\geq 2$, and we will then attempt to prove them for $n$. 
We being with the calculation of $\delta^{n-1}e^{isK/2\pi}$. From equation \eqref{expder}, we get by taking $n-2$ more derivatives:
\beqn
\delta^{n-1} e^{\frac{isK}{2\pi}}&=& \frac{i}{2\pi}\sum_{m_1+m_2+m_3= n-2}\frac{(n-2)!}{m_1!m_2!m_3!}\int_0^sdt\,(\delta^{m_1}e^{\frac{itK}{2\pi}})(\delta^{m_2+1}K) (\delta^{m_3}e^{\frac{i(s-t)K}{2\pi}})\\
&=& -\frac{(-i)^{n-1}}{(2\pi)^{n-1}}(n-2)!\sum_{m_1+m_2+m_3= n-2}(m_2+1)\int d\mu_{n-1}\int_0^sdt\int_{i\tau_1} ds_1\cdots \int_{i\tau_{n-1}}ds_{n-1}\nonumber\\
&\times &4\sinh^2(t/2)f_{m_1+1}(s_1,\cdots, s_{m_1},t)O_1(s_1)\cdots O_{m_1}(s_{m_1})e^{\frac{itK}{2\pi}}\nonumber\\
&\times & f_{m_2+1}(s_{m_1+1},\cdots, s_{m_1+m_2+1})O_{m_1+1}(s_{m_1+1})\cdots O_{m_1+m_2+1}( s_{m_1+m_2+1})\nonumber\\
&\times &4\sinh^2((s-t)/2)f_{m_3+1}(s_{m_1+m_2+2},\cdots, s_{n-1},s-t)O_{m_1+m_2+2}(s_{m_1+m_2+2})\cdots O_{n-1}(s_{n-1})e^{\frac{i(s-t)K}{2\pi}},\nonumber
%&=& -\frac{i^{n}}{(2\pi)^n}(n-1)!\int ds_1\cdots ds_n O(s_1)\cdots O(s_n)2\sinh(s/2)f_{(n+1)}(s_1,\cdots s_n, s)\\
%&\times &\sum_{m_1+m_2+m_3= n-1}(m_2+1)\int_0^sdt\,\frac{\sinh(t/2)\sinh(\frac{s-t}{2})\sinh(\frac{s_{m_1+1}-s_{m_1}}{2})\sinh(\frac{s_{m_1+m_2+2}-s_{m_1+m_2+1}}{2})}{\sinh(\frac{t-s_{m_1}}{2})\sinh(\frac{s_{m_1+1}-t}{2})\sinh(\frac{s_{m_1+m_2+1}-t}{2})\sinh(\frac{s_{m_1+m_2+2}-t}{2})}e^{isK}\nonumber
\eeqn
where for brevity we have denoted $O_i(s_i) = e^{\frac{is_i K}{2\pi}}O(\tau_i,Y_i)e^{-\frac{is_i K}{2\pi}}$. We can simplify this expression by using the identity
\beq
\frac{\sinh(t/2)\sinh\left(\frac{s_2-s_1}{2}\right)}{\sinh\left(\frac{s_2-t}{2}\right)\sinh\left(\frac{s_1-t}{2}\right)}=\frac{\sinh(s_1/2)}{\sinh\left(\frac{s_1-t}{2}\right)}-\frac{\sinh(s_2/2)}{\sinh\left(\frac{s_2-t}{2}\right)}.
\eeq
So we get
\beqn
\delta^{n-1} e^{\frac{isK}{2\pi}}&=&-\frac{(-i)^{n-1}}{(2\pi)^{n-1}}(n-2)!\int d\mu_{n-1}\int_{i\tau_1} ds_1\cdots \int_{i\tau_{n-1}}ds_{n-1} O_1(s_1)\cdots O_{n-1}(s_{n-1})2\sinh(s/2)\nonumber\\
&\times &f_{(n)}(s_1,\cdots s_{n-1}, s)\int_0^sdt \sum_{m_1+m_2+m_3= n-2}(m_2+1)\left[\frac{\sinh(s_{m_1+1}/2)}{\sinh\left(\frac{s_{m_1+1}-t}{2}\right)}-\frac{\sinh(s_{m_1}/2)}{\sinh\left(\frac{s_{m_1}-t}{2}\right)}\right]\nonumber\\
&\times &\left[\frac{\sinh(\frac{s_{n-m_3}-s}{2})}{\sinh\left(\frac{s_{n-m_3}-t}{2}\right)}-\frac{\sinh(\frac{s_{n-m_3-1}-s}{2})}{\sinh\left(\frac{s_{n-m_3-1}-t}{2}\right)}\right]e^{\frac{isK}{2\pi}}.
\eeqn
Now it's a simple matter to perform the sum, and we find
\beqn
\delta^{n-1} e^{isK}&=&\frac{(-i)^{n-1}}{(2\pi)^{n-1}}(n-2)!\int d\mu_{n-1}\int_{i\tau_1} ds_1\cdots \int_{i\tau_{n-1}}ds_{n-1} O_1(s_1)\cdots O_n(s_{n-1})\nonumber\\ &\times &2\sinh(s/2)f_{(n)}(s_1,\cdots s_{n-1}, s)\sum_{j=1}^{n-1}\int_0^sdt\,\frac{\sinh(s_j/2)\sinh(\frac{s_j-s}{2})}{\sinh^2(\frac{s_j-t}{2})}e^{\frac{isK}{2\pi}}\\
&=&\frac{(-i)^{n-1}}{(2\pi)^{n-1}}(n-1)!\int d\mu_{n-1}\int_{i\tau_1} ds_1\cdots \int_{i\tau_{n-1}}ds_{n-1} O_1(s_1)\cdots O_{n-1}(s_{n-1})\frac{f_{(n)}(s_1,\cdots s_{n-1}, s)}{f_{(1)}(s)} e^{\frac{isK}{2\pi}}.\nonumber
\eeqn

The next step is to calculate $\delta^n K $. Once again, we start by taking $(n-1)$ derivatives of \eqref{exp1}:
\beq 
\delta^n K = \delta^{n-1}\left( \int_{-\infty + i \epsilon}^{\infty + i\epsilon} ds~ f_{(1)}(s)~ e^{\frac{i s K }{2\pi} } \left(\rho^{-1} \delta \rho \right) e^{-\frac{i s K }{2\pi} } \right) = \sum_{b=0}^{n-1} T(b),
\eeq
where $T(b)$ denotes the terms with $\delta^b $ acting on $\left(\rho^{-1}\delta \rho\right)$. We have chosen the contour to be (almost) real so that we can apply the previously derived formulae for the derivatives of $e^{\frac{i s K }{2\pi} }$.  
We introduce some further notation for the sake of brevity of expressions. Let 
$$ \tilde{O}_p  \equiv e^{\frac{i s_p K }{2\pi} } O(\tau_p , Y_p) e^{- \frac{i s_p K }{2\pi} }, $$
and whenever $\tilde{O}_p$ appears in expressions, the  integrals $\int d\tau_p \int dY_p ~\lambda(\tau_p,Y_p) $ sitting outside the expression is implicit.  Also note that when $\text{Im}(s_p) = \tau_p$, the operator $\tilde{O}_p $ is at a purely Lorentzian point.
\begin{align}
T(0) &=  \sum_{j=0}^{n-1} \int ds_p ~ f_{(1)}(s_p) \binom{n-1}{j} \left( \delta^{n-1-j} e^{\frac{i s_p K }{2\pi} } \right)  \left(\rho^{-1} \delta \rho \right) \left( \delta^j  e^{\frac{- i s_p K }{2\pi} } \right)  \\
&= (n-1)!\left(\frac{(-i)}{2\pi}\right)^{n-1}~ \sum_{j=0}^{n-1} \left(\prod_{q \neq n-j} \int_{\mathbb{R}+i \tau_q} d s_q\right) \int_{\mathbb{R} + i \epsilon}  d s_{n-j}   \nonumber \\
&~~~~~\times 
4(-1)^j  \sinh^2 s_{n-j}~  f_{(n-j)}(s_1,s_2,\ldots s_{n-j}) f_{(j+1)}(s_n,s_{n-1},\ldots s_{n-j})
\tilde{O}_1 \tilde{O}_2 \ldots \tilde{O}_n \\
&= (n-1)!\left(\frac{(-i)}{2\pi}\right)^{n-1}~ \sum_{j=0}^{n-1} \left(\prod_{q \neq n-j} \int_{\mathbb{R}+i \tau_q} d s_q\right) \left(\int_{\mathbb{R} + i \epsilon}  d s_{n-j} \right) f_{(n)}(s_1,s_2,\ldots s_n) \tilde{O}_1 \tilde{O}_2 \ldots \tilde{O}_n \label{eq:t0} \\
&\equiv Q + P(0)
\end{align}
where $Q$ is the expression we would get by lifting the $s_{n-j}$ contour from $\mathbb{R}+i\epsilon$ to $\mathbb{R}+i\tau_{n-j}$ in the $j$th term in  (\ref{eq:t0}), and $P(0)$ is what we will call a pole contribution because these terms come from the poles which were crossed in this contour deformation. More explicitly, we have
\beq
Q = (n-1)!\left(\frac{(-i)}{2\pi}\right)^{n-1}~ \sum_{j=0}^{n-1} \left(\prod_{q=1}^n \int_{\mathbb{R}+i \tau_q} d s_q\right) f_{(n)}(s_1,s_2,\ldots s_n) \tilde{O}_1 {\tilde{O}_2} \ldots \tilde{O}_n .
\label{eq:Q}
\eeq
Note that the sum over $j$ above simply gives a factor of $n$, and also that all the operators are inserted in a purely Lorentzian kinematic regime.  Surprisingly, we will find that the pole contributions cancel with the remaining terms:
\beq 
P(0) + \sum_{b=1}^{n-1} T(b) = 0,
\label{eq:cancellation}
\eeq
so that $\delta^n K =  Q$ thus proving our claim equation \eqref{induction2}, which remarkably is a purely Lorentzian formula for $\delta^n K $. So the surprising cancellation in \eqref{eq:cancellation} eventually leads us to this rather simple formula. We now turn to showing these cancellations.  

\subsection{Cascading pole cancellations}
In the following expressions, we will use the abbreviation $\int_{\{s\}}$ to represent list of $s$ integrals which are to be understood by looking at the expression in the integrand. Further, the contour of $s_i$ is understood to be at $\mathbb{R} + i \tau_i$ unless otherwise specified.
\begin{align}
P(0) &= T(0)-Q \\
&= (-2 \pi i)  \left( \frac{(-i)}{2\pi} \right) ^{n-1} (n-1)! \sum_{j=0}^{n-1} \int_{\{s\}} f_{(n-1)} (s_1,\ldots s_{n-j-1},s_{n-j+1},\ldots s_n) \nonumber \\
& ~~~~~\times \Bigg( \Theta(\tau_{n-j} - \tau_{n-j+1}) \tilde{O}_1\ldots \tilde{O}_{n-j-1} (O_{n-j}O_{n-j+1})_{s_{n-j+1} } \tilde{O}_{n-j+2} \ldots \tilde{O}_n  \nonumber \\ 
& ~~~~~~~~~~ - \Theta(\tau_{n-j} - \tau_{n-j-1} ) \tilde{O}_1 \ldots \tilde{O}_{n-j-2} (O_{n-j-1}O_{n-j})_{s_{n-j-1} } \tilde{O}_{n-j+1}\ldots \tilde{O}_{n}  \Bigg)
\label{eq:P0temp1}
\end{align}
In (\ref{eq:P0temp1}) we have plugged in for $T(0)$ and $Q$ and performed $s_{n-j}$ integral in the $j$th term via residue integration (and one can close the contour at infinity due to the exponential fall off). The $\Theta$'s in (\ref{eq:P0temp1}) appear in order to keep track of whether or not the corresponding pole is contained in the contour. To clarify notation, the $\int_{\{s\} } $ in (\ref{eq:P0temp1}) stands for $(n-1)$ $s$-integrals, i.e., $\int ds_1 \ldots \int ds_{n-j-1} \int ds_{n-j+1} \ldots \int ds_{n}$.  Also note that the first term in (\ref{eq:P0temp1}) should be set to $0$ for $j=0$ and the second term should be set to $0$ for $j=n-1$, which can be achieved by setting $\tau_0= \tau_{n+1} = 2\pi$ (say). Further $O_j$ (without tilde) is a shorthand for $O(\tau_j, Y_j)$ and $(X)_s$ is a shorthand  for $e^{\frac{i s K }{2\pi}} X e^{-\frac{i s K }{2\pi}} $. 

Replace $j$ with $j+1$ in the first term (i.e., do $j=j'+1$ first followed by $j'=j$). Then the sum can be restricted to go  from $j=0,1,\ldots (n-2)$. Doing so, the two terms in  (\ref{eq:P0temp1}) read
\begin{align}
P(0)|_{\rni} &= -(2 \pi i)  \left( \frac{(-i)}{2\pi} \right) ^{n-1} (n-1)! \sum_{j=0}^{n-2} \int_{\{s\}} f_{(n-1)}(s_1,\ldots,s_{n-j-2},s_{n-j},\ldots s_n) \nonumber\\ &~~~~~~~~\times\Theta(\tau_{n-j-1}-\tau_{n-j}) \tilde{O}_1 \ldots \tilde{O}_{n-j-2} (O_{n-j-1}O_{n-j})_{s_{n-j} } \tilde{O}_{n-j+1}\ldots \tilde{O}_{n} \\
P(0)|_{\rnii } &= (2 \pi i)  \left( \frac{(-i)}{2\pi} \right) ^{n-1} (n-1)! \sum_{j=0}^{n-2} \int_{\{s\}} f_{(n-1)}(s_1,\ldots,s_{n-j-1},s_{n-j+1},\ldots s_n) \nonumber\\ &~~~~~~~~\times\Theta(\tau_{n-j}-\tau_{n-j-1}) \tilde{O}_1 \ldots \tilde{O}_{n-j-2} (O_{n-j-1}O_{n-j})_{s_{n-j-1} } \tilde{O}_{n-j+1}\ldots \tilde{O}_{n} ,
\end{align}
where $|_{\rni}$ denotes first term and $|_{\rnii}$ denotes second term. (Henceforth, to be systematic, when computing $s_i$ contour integral, the pole at $s_k$ with $k>i$ will be labelled as ``first term'').
It is tempting to relabel $s_{n-j}$ in the first term to $s_{n-j-1}$ so that the two terms can be combined. However this is not naively allowed since $\text{Im}(s_{n-j}) = \tau_{n-j}$ and $\text{Im}(s_{n-j-1}) = \tau_{n-j-1}$.  Hence we push down the $s_{n-j}$ contour in the first term down to imaginary part $\epsilon$ and do the same with the $s_{n-j-1}$ contour  in the second term (and then relabel $s_{n-j}$ with $s_{n-j-1}$ in the first term, which we now can since both contours are identical). We have to add ``pole-contributions'' that account for the poles crossed while pushing the contour down, i.e.,
\beq P(0) = \tilde{P}(0) + P(1),\eeq
where $\tilde{P}(0)$ is the term obtained by doing the contour manipulations on $P(0)$ explained above and $P(1)$ is the ``pole contribution''.  Before we compute $P(1)$, we first show that $\tilde{P}(0)+T(1) = 0$. To do so, we start by writing 
\begin{align}
T(1) &= \sum_{j=0}^{n-2} \int_{\mathbb{R}+i\epsilon} ds_p~ f_{(1)}(s_p) \binom{n-2}{j} (n-1) \left( \delta^{n-2-j} e^{\frac{i s_p K }{2\pi}}\right)  \left( \rho^{-1} \delta^2 \rho - (\rho^{-1} \delta \rho)^2 \right) \left( \delta^{j} e^{- \frac{i s_p K }{2\pi}} \right)  \\
&=
\sum_{j=0}^{n-2}~~\int_{\{s\}} \left(\frac{(-i)}{2\pi} \right)^{n-2} (n-1)!  f_{(n-1)} (s_1,\ldots s_{n-j-1},s_{n-j+1},\ldots s_n)   \nonumber \\ 
&~~~~~~~~~~\times \big(2\theta(\tau_{n-j-1}-\tau_{n-j})-1\big) \tilde{O}_1 \ldots \tilde{O}_{n-j-2} \left( O_{n-j-1}O_{n-j} \right)_{s_{n-j-1}} \tilde{O}_{n-j+1}\ldots \tilde{O}_n,
\label{eq:T1}
\end{align}
where the $s_{n-j-1}$-integral in the $j$th term in (\ref{eq:T1}) is at  $\text{Im}(s_{n-j-1}) = \epsilon$. Then we have
\begin{align}
\tilde{P}(0) + T(1) &= \sum_{j=0}^{n-2} \int_{\{ s \} } (n-1)! \left(\frac{-i}{2\pi} \right)^{n-2} f_{(n-1)}(s_1,\ldots s_{n-j-1},s_{n-j+1},\ldots s_n) \nonumber \\
&~~~~\times \left( 2 \Theta(\tau_{n-j-1}-\tau_{n-j} ) - 1 - \Theta(\tau_{n-j-1}-\tau_{n-j}) + \Theta(\tau_{n-j}-\tau_{n-j-1}) \right) \label{eq:cancel0}
\\
&=0.
\end{align}
Note that in the above sum, each term vanishes individually since this particular combination of $\Theta$'s vanishes, which is easily verified by using the identity  $ \Theta(\tau_j-\tau_i) = 1- \Theta(\tau_i-\tau_j) $. 
We then have
\beq P(0) + T(1) = P(1) \equiv P(1)|_{\rni} + P(1)|_{\rnii}, \eeq
where
\begin{align}
P(1)|_{\rni} &= 
(-2\pi i)^2 \left(\frac{(-i)}{2\pi}\right)^{n-1}(n-1)! \sum_{j=0}^{n-2} \int_{\{s\} } f_{(n-2)} (s_1,\ldots,s_{n-j-2},s_{n-j+1},\ldots,s_n)  \nonumber \\
&~~~~~~\times \Theta(\tau_{n-j-1}-\tau_{n-j}) \Bigg(
 - \Theta(\tau_{n-j}-\tau_{n-j+1}) \tilde{O}_1^{n-j-2} \left( O_{n-j-1}^{n-j+1} \right)_{s_{n-j+1} } \tilde{O}_{n-j+2}^{n} \nonumber \\
 &~~~~~~~~~~~~~~~~~~~~~~~~~~~+\Theta(\tau_{n-j}-\tau_{n-j-2})\tilde{O}_1^{n-j-3}\left( O_{n-j-2}^{n-j} \right)_{s_{n-j-2}} \tilde{O}_{n-j+1}^n \Bigg) \label{eq:P1_1}\\
P(1)|_{\rnii} &= (-2\pi i)^2 \left(\frac{(-i)}{2\pi}\right)^{n-1}(n-1)! \sum_{j=0}^{n-2} \int_{\{s\} } f_{(n-2)}(s_1,\ldots s_{n-j-2},s_{n-j},\ldots s_n)  \nonumber \\
&~~~~~~\times \Theta(\tau_{n-j}-\tau_{n-j-1})  \Bigg(
 \Theta(\tau_{n-j-1} - \tau_{n-j+1} ) \tilde{O}_1^{n-j-2} \left( O_{n-j-1}^{n-j+1}\right)_{s_{n-j+1}} O_{n-j+2}^{n} \nonumber \\
&~~~~~~~~~~~~~~~~~~~~-\Theta(\tau_{n-j-1}-\tau_{n-j-2}) \tilde{O}_1^{n-j-3} \left( O_{n-j-2}^{n-j} \right)_{s_{n-j-2}} \tilde{O}_{n-j+1}^{n} \Bigg),
\label{eq:P1_2}
\end{align}
where we have introduced the shorthand $X_i^j \equiv \prod_{k=i}^{j} X_k$, for $X=O,\tilde{O}$.
The above expressions have been obtained by writing $P(1) \equiv P(0) - \tilde{P}(0)$ and doing the $s_{n-j}$ contour  integral in first term and $s_{n-j-1}$ contour integral in second term. 
The definitions $\tau_0=\tau_{n+1}= 2\pi$ will continue to take care of corner case terms which are required to vanish. As an aside, note that at $n=2$, we are done with the proof, since $P(1)$ consists of only the corner case terms and vanishes.  

Now, replace $j$ with $j+1$ in the first terms of (\ref{eq:P1_1}) and (\ref{eq:P1_2}). This will allow us to change the range of summation to $j=0 \ldots (n-3)$.  Before we write things down, we introduce a useful shorthand: 
\beq
\Theta({\vec{\tau}}) \equiv \prod_{i=1}^{\text{length}(\vec{\tau})-1 } \Theta(\tau[i] - \tau[i+1]). 
\eeq
Further we write,
\beq P(1) = \sum_{( A_1,A_2) \in \{\rni,\rnii \}^2 } P(1)|_{( A_1,A_2 ) }, \eeq where $P(1)_{ (A_1, A_2 ) } $ is the ${A}_2$th term in $P(1)|_{{A}_1}$ computed earlier in (\ref{eq:P1_1}),(\ref{eq:P1_2}). We find
\begin{align}
P(1)|_{(\rni,\rni)} 
 &= (-2\pi i )^2 \left( \frac{(-i)}{2\pi}\right)^{n-1} (n-1)! \sum_{j=0}^{n-3} \int_{\{s \} } f_{(n-2)} (s_1,\ldots,s_{n-j-3}, s_{n-j},\ldots,s_n) \nonumber \\ &~~~~~~~~~~\times \left(-\Theta((\tau_{n-j-2},\tau_{n-j-1},\tau_{n-j}) ) \right) \tilde{O}_1^{n-j-3} \left( O_{n-j-2}^{n-j} \right)_{s_{n-j}} \tilde{O}_{n-j+1}^{n}  \Bigg)  \\
P(1)|_{(\rni,\rnii )}  &= (-2\pi i )^2 \left( \frac{(-i)}{2\pi} \right)^{n-1} (n-1)! \sum_{j=0}^{n-3} \int_{\{s\} } f_{(n-2)}(s_1,\ldots,s_{n-j-2},s_{n-j+1},\ldots,s_n) \nonumber \\
&~~~~~~~~~~\times  \Theta((\tau_{n-j-1},\tau_{n-j},\tau_{n-j-2})) \tilde{O}_1^{n-j-3} \left( O_{n-j-2}^{n-j} \right)_{s_{n-j-2}} \tilde{O}_{n-j+1}^{n}  
\end{align}

\begin{align}
P(1)|_{(\rnii,\rni) } &= (-2\pi i)^2 \left(\frac{(-i)}{2\pi}\right)^{n-1}(n-1)! \sum_{j=0}^{n-3} \int_{\{s\} } f_{(n-2)}(s_1,\ldots s_{n-j-3},s_{n-j-1},\ldots s_n)  \nonumber \\
&~~~~~~\times  \left(- \Theta((\tau_{n-j-1}, \tau_{n-j-2},\tau_{n-j} )) \right) \tilde{O}_1^{n-j-3} \left( O_{n-j-2}^{n-j}\right)_{s_{n-j}} \tilde{O}_{n-j+1}^{n} \\
P(1)|_{(\rnii ,\rnii) } &= (-2\pi i)^2 \left(\frac{(-i)}{2\pi}\right)^{n-1}(n-1)! \sum_{j=0}^{n-3} \int_{\{s\} }
f_{(n-2)}(s_1,\ldots s_{n-j-2},s_{n-j},\ldots s_n) \nonumber \\ &~~~~~~\times \Theta_{{n-j},{n-j-1}}
\Theta_{{n-j-1},{n-j-2}} \tilde{O}_1^{n-j-3} \left( O_{n-j-2}^{n-j} \right)_{s_{n-j-2}} \tilde{O}_{n-j+1}^{n} 
\end{align}
Like in the case of $P(0)$,  we wish to combine terms in this expression after relabelling particular $s$-variables. But the contour prescription doesn't naively allow the relabelling, and we write
\beq
P(1)|_{\alpha} = \tilde{P}(1)|_{\alpha} + P(2)|_{\alpha},
\eeq
where $\alpha \in \{\rni,\rnii\}^2 $ and $\tilde{P}(1)|_{\alpha}$ represents the quantity with a particular $s$ variable's contour changed to imaginary part $\epsilon$. For instance, when $\alpha=(\rni,\rni) \text{ and } (\rnii,\rni)$, we modify the $s_{n-j}$ contour to go down to $\epsilon$. For $\alpha = (\rni, \rnii) \text{ and  } (\rnii,\rnii)$,  we modify $s_{n-j-2}$ contour down to $\epsilon$. 
After doing so, we then have to perform allowed relabelling of $s$-variables  to show that $\tilde{P}(1)+T(2) = 0$. Before doing so, we first generalize this algorithm. We write
\beq P(b-1) = \sum_{\alpha\in \{\rni,\rnii \}^b } P(b-1)|_\alpha  ,\eeq
where
\begin{align}
P(b-1)|_\alpha &= \sum_{j=0}^{n-b-1} \gamma_j(b-1) f_{(n-b)}(s_1,\ldots s_{n-j-b-1},  s({i_\alpha^{n,j}}), s_{n-j+1}\ldots s_n) (-1)^{\sigma_\alpha}~ 
\Theta\left(\vec{\tau}_\alpha^{~n,j}\right)  \nonumber \\
&~~~~~~~~~~~\times \tilde{O}_{1}^{n-j-b-1} \left(O_{n-j-b}^{n-j} \right)_{s({i_\alpha^{n,j}) } } \tilde{O}_{n-j+1}^{n}
\label{eq:Pb}
\end{align}
where  $\gamma_j(b-1) = (-2\pi i )^b \left(\frac{(-i)}{2\pi} \right)^{n-1} (n-1)! $ , and  $\vec{\tau}_{\alpha}^{~n,j}$ is an $\alpha$-dependent  permutation of $(\tau_{n-j-b},\ldots, \tau_{n-j} )$,  and $s(i_\alpha^{n,j}) \equiv s_{i_\alpha^{n,j} }$ is the particular variable whose contour needs to be pushed down to $\epsilon$.  These quantities are recursively defined as follows
\beq
\sigma_{(\alpha \rni)} = \sigma_\alpha+1, ~~~~~~ \sigma_{(\alpha \rnii)}= \sigma_\alpha
\eeq
\beq
i_{(\alpha \rni)}^{n,j} = n-j,  ~~~~~~ i_{(\alpha \rnii)}^{n,j} = n - j - \text{Length}(\alpha) - 1
\eeq
\beq
\vec{\tau}_{(\alpha \rni)}^{~n,j} = (\vec{\tau}_\alpha^{~n,j+1} , \tau_{n-j} ), ~~~~~~~~~~ 
\vec{\tau}_{(\alpha \rnii)}^{~n,j} = (\vec{\tau}_\alpha^{~n,j} , \tau_{n-j-\text{Length}(\alpha)-1} ).
\label{eq:taupermutation}
\eeq
The base cases of the definitions of these quantities can be gleaned from the formula for $P(0)$. 

The contour modification and relabelling step involves modifying the  $s_{i_\alpha^{n,j} }$ down to imaginary part $\epsilon$   in the $j$th term in the sum for $P(b-1)|_\alpha$ and relabelling it  to $s$(say), to compute $\tilde{P}(b-1)$. The next step in the algorithm is to compute $\tilde{P}(b-1) + T(b)$ and show that it vanishes. These steps (applied repeatedly) are sufficient to prove (\ref{eq:cancellation}) (since eventually we will be left with an empty sum for $P(n-1)$).
We first compute $T(b)$.
\begin{align}
T(b) &=  \int_{\mathbb{R}+i\epsilon} ds_p~ f_{(1)}(s_p) \sum_{j=0}^{n-b-1} \frac{ (n-1)!}{b! j! (n-b-j-1)!} \left(\delta^{n-b-1-j}e^{\frac{i s_p K}{2\pi} }\right) (\delta^b(\rho^{-1} \delta \rho)) \left(\delta^j e^{-\frac{i s_p K}{2\pi} }\right) \\
&=  \left(\frac{(-i)}{2\pi}\right)^{n-b-1}\frac{(n-1)!}{b!} \sum_{j=0}^{n-b-1}   \int_{\mathbb{R}+i\epsilon} ds_{n-b-j} \int_{\{s\} } f_{(n-b)}(s_1,\ldots,s_{n-b-j},s_{n-j+1},\ldots,s_n) \nonumber \\
&~~~~~~~~~~~~~~~~~~\times \tilde{O}_1 \ldots \tilde{O}_{n-b-j-1} \left(\delta^{b}(\rho^{-1}\delta\rho)\right)_{s_{n-b-j}} \tilde{O}_{n-j+1}\ldots \tilde{O}_{n}.
\label{eq:Tb}
\end{align}
We need to  compute $\delta^{b}(\rho^{-1}\delta\rho)$, which can be systematized as follows. Let $\mathcal{P}_{b+1}$ be the set of ordered partitions of $b+1$. Note that there are $2^b$ ordered partitions of $b+1$. These ordered partitions can be put into a one-to-one correspondence with the numbers from $2^b$ to $2^{b+1}-1$ when viewed as binary bit strings. Also, to an ordered partition $p$, we define $\mathbf{\Theta}_p(\tau_1,\tau_2,\ldots, \tau_{b+1}) $ to be a $p$-dependent product of $\Theta$'s. We will illustrate the definitions of $\mathbf{\Theta}$ with  the example of the ordered partition of $8$ given by $(2,3,1,2)$. The bit-string\footnote{To an ordered partition $p\equiv(p_1,p_2,\ldots,p_k)$ satisfying $\sum_{i=1}^k p_i = b+1$,  we  associate a $b+1$ length bit string beginning with `1' which is constructed as follows: `1' followed by $(p_1-1)$ `0's, followed by `1' followed by $(p_2-1)$ `0's , \ldots, followed by `1` followed by $(p_k - 1) $ `0's. This is a one-to-one and invertible map between ordered partitions and bit strings.} associated to which is ``10100110" which corresponds to the decimal integer\footnote{This remark is just to indicate how it can be encoded as an integer for the purposes of a computer program.} $166$. The product of $\Theta$'s associated with which is: $$\mathbf{\Theta}_{10100110}(\tau_1,\tau_2,\ldots, \tau_8) = \Theta(\tau_1-\tau_2)\Theta(\tau_3-\tau_4)\Theta(\tau_4-\tau_5)\Theta(\tau_7-\tau_8)$$
i.e., for the ordered partition $(2,3,1,2)$, $\mathbf{\Theta}_{10100110}$ will be used to require that $(1,2)$ be time ordered, $(3,4,5)$ be time ordered, $(7,8)$ are time ordered. Henceforth we will use ordered partitions, bit strings of length $b+1$ which begin with $1$, an integer between $2^b$ and $2^{b+1}-1$ interchangeably. (We will mostly use bit strings notation). We will also define a numerical coefficient $\mathbf{n}(p)$ which is simply the product of factorials of the parts. i.e. for the example we considered, $\mathbf{n}(10100110) = 2!3!1!2! = 24$.

We have
\beq
\delta^b (\rho^{-1} \delta \rho) = \left(\sum_{p \in \mathcal{P}_{b+1} } c(p) \mathbf{n}(p) \mathbf{\Theta}_p(\tau_{n-b-j},\ldots \tau_{n-j}) \right) O_{n-b-j}\ldots O_{n-j}
\label{eq:bterm}
\eeq
where the numerical coefficients $c(p)$ satisfy the recursion relation (which simply follows from the product rule of differentiation)
\beq
c(p) = \sum_{p' \in \mathcal{P}_+(p) } c(p') - \sum_{p' \in \mathcal{P}_-(p) } c(p')
\label{eq:crecursion}
\eeq
 The list $\mathcal{P}_+(p)$ corresponds to a list of bit strings that can be obtained from $p$ by deleting a single  occurence of $0$ bit that is either the last bit or an occurence of a $0$ bit that is immediately followed by a $1$ bit.
 The list $\mathcal{P}_-(p)$ corresponds to a list of bit strings that can be obtained from $p$ by deleting a single occurence of a $1$ bit which is immediately followed by another occurence of $1$ bit. We call these objects lists (as opposed to sets), because they are allowed to have repeated entries as will be illustrated shortly. We remind ourselves that there is an implicit integration over the spacetime points (together with appropriate $\lambda$'s)  in the RHS of (\ref{eq:bterm}).  We now illustrate the above recursion  for some low values of $b$ to clarify definitions.
\begin{itemize}
\item[$b=0$:] For $b=0$, we have only one ordered partition of $b+1=1$ whose bit string is simply $1$. We define $c(1) = 1$. This is the base case.
\item[$b=1$:] We have two two-bit strings: $\{10, 11\}$. Following the rules, we find 
$\mathcal{P}_+(10) = \{1\}, ~\mathcal{P}_-(11) = \{1\} $.
Hence we have $c(10) = c(1) = 1$, and , $c(11)= -c(1) = -1$.
This basically corresponds to the fact that $\delta(\rho^{-1} \delta\rho) = -\rho^{-1} \delta\rho \rho^{-1} \delta\rho+\rho^{-1} \delta^2\rho$. 
Also, we have $\mathbf{\Theta}_{11}(\tau_1,\tau_2) = 1, \mathbf{\Theta}_{10}(\tau_1,\tau_2) = \Theta(\tau_1-\tau_2)$.  Plugging this into (\ref{eq:bterm}) we get
$$\delta(\rho^{-1} \delta\rho) = \big(2!\Theta(\tau_1-\tau_2)-1\big)O_1 O_2. $$
\item[$b=2$:]We have $4$ bit strings: $\{100,101,110,111\}$. Following the rules, we get
$$ \mathcal{P}_+(100)=\{10\}, ~~ \mathcal{P}_+(101)=\{11\},~~ \mathcal{P}_+(110)=\{ 11\},~~
\mathcal{P}_+(111)=\{\}
$$
$$ \mathcal{P}_-(100)=\{\}, ~~ \mathcal{P}_-(101)=\{\},~~ \mathcal{P}_-(110)=\{10\},~~
\mathcal{P}_-(111)=\{11,11\}
$$
Note that $P_-(p)$ can have repeated entries. 
This data above corresponds to 
$$c(100)=c(10)=1,~~ c(101)=c(11)=-1,~~ c(110)=c(11)-c(10)=-2,~~c(111)=-2 c(11)=2 .$$
This is in agreement with 
 $\delta^2 (\rho^{-1}\delta\rho) =  \rho^{-1} \delta^3\rho - (\rho^{-1} \delta^2\rho)(\rho^{-1} \delta\rho) -2 (\rho^{-1} \delta\rho)(\rho^{-1} \delta^2\rho) +2 (\rho^{-1} \delta\rho)^3 .$
\end{itemize}  
Putting (\ref{eq:Pb}),(\ref{eq:Tb}),(\ref{eq:bterm}) together we deduce that $\tilde{P}_{b-1} + T(b) = 0$ if the following purely combinatorial identity holds:
\beq
\left( \sum_{\alpha \in \{\rni,\rnii \}^b } (-1)^{\sigma_\alpha} \Theta\left(\vec{\tau}_\alpha^{~n,j}  \right) \right) + \left( \sum_{p \in P_{b+1}} \frac{c(p)\mathbf{n}(p) }{b!} \mathbf{\Theta}_p(\tau_{n-b-j},\ldots, \tau_{n-j}) \right) = 0.
\label{eq:cancel}
\eeq
A general proof of this identity for arbitrary values of $b$ is given in Appendix \ref{app:identity}. Additionally, we have verified this identity for $b=1,2,\ldots 8$  using a computer program utilizing the recursive definitions of the quantities that appear in it. As argued earlier, checking (\ref{eq:cancel}) for $1\leq b\leq b_*$ proves our formula for $\delta^n K$ for $n = 1,2,\ldots b_* + 1 $. \footnote{In practice, computer-aided checking is limited by the fact that the computational time complexity of the naive checking of  (\ref{eq:cancel}) for a given $n$ is  is $\text{O}(n!~ 2^n)$, where  O here stands for the big-O notation of computational complexity. This naive checking involves verifying that (\ref{eq:cancel}) holds for all possible orderings of the $\tau$'s. The number of orderings is O$(n!)$ and the number of terms in the identity is O$(2^n)$. } This completes our proof.  

\section{Modular Hamiltonians for Shape Deformed Half-spaces}\label{sec:sec3}
In the previous section, we derived the modular Hamiltonian for excited states created by turning on Euclidean path-integral sources for local operators, with the subregion being a Rindler wedge/half-space. In this section, we wish to apply this formula to derive the modular Hamiltonian for the vacuum state, but where the subregion is a deformed half-space (see \cite{Faulkner:2016mzt, Rosenhaus:2014woa, Rosenhaus:2014zza, Allais:2014ata, Mezei:2014zla, Faulkner:2015csl, Dong:2016wcf, Bianchi:2016xvf} for some previous work on shape deformations of entanglement/Renyi entropy). 

\begin{figure}
    \centering
      \includegraphics[height=5cm]{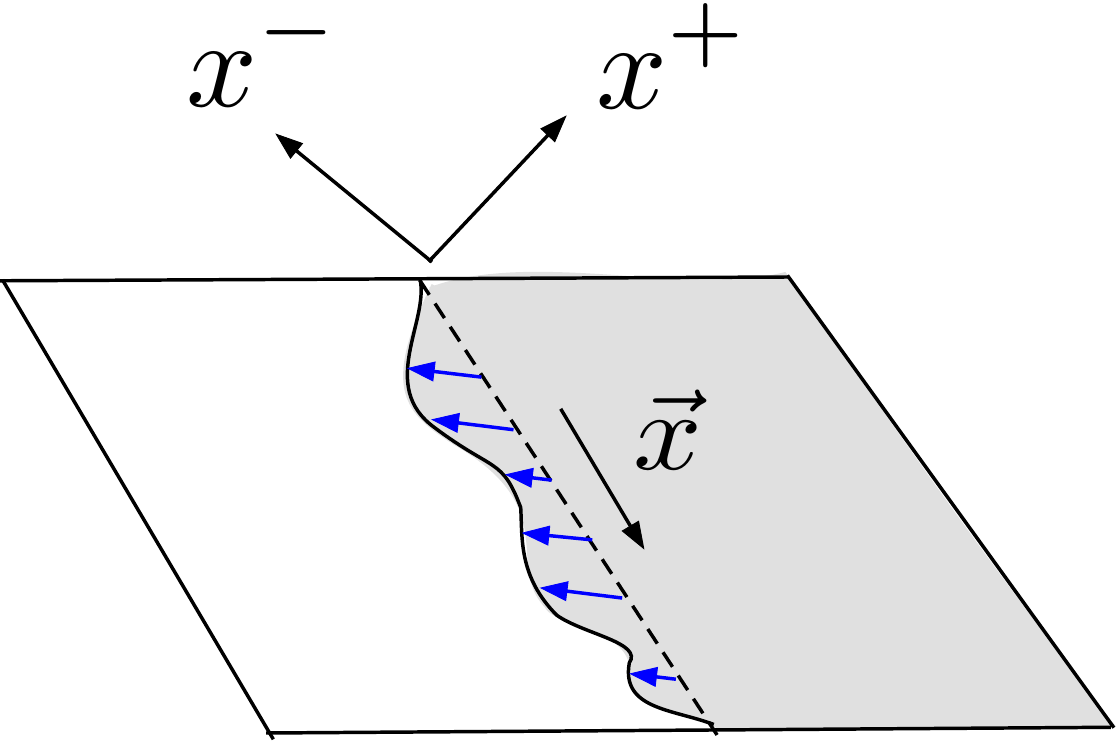} 
    \caption{\small{The plane denotes the a constant time Cauchy slice. We deform the shape of the subregion from the original half-space (dashed line) to the shaded region. The blue arrows denote the diffeomorphism $V=(V^+(\vec{x}), V^-(\vec{x}), \vec{0})$ which maps the old entanglement cut to the new one.}}
    \label{fig:shape}
\end{figure}

\subsection{General shape deformations}
\begin{figure}
    \centering
      \includegraphics[height=5.5cm]{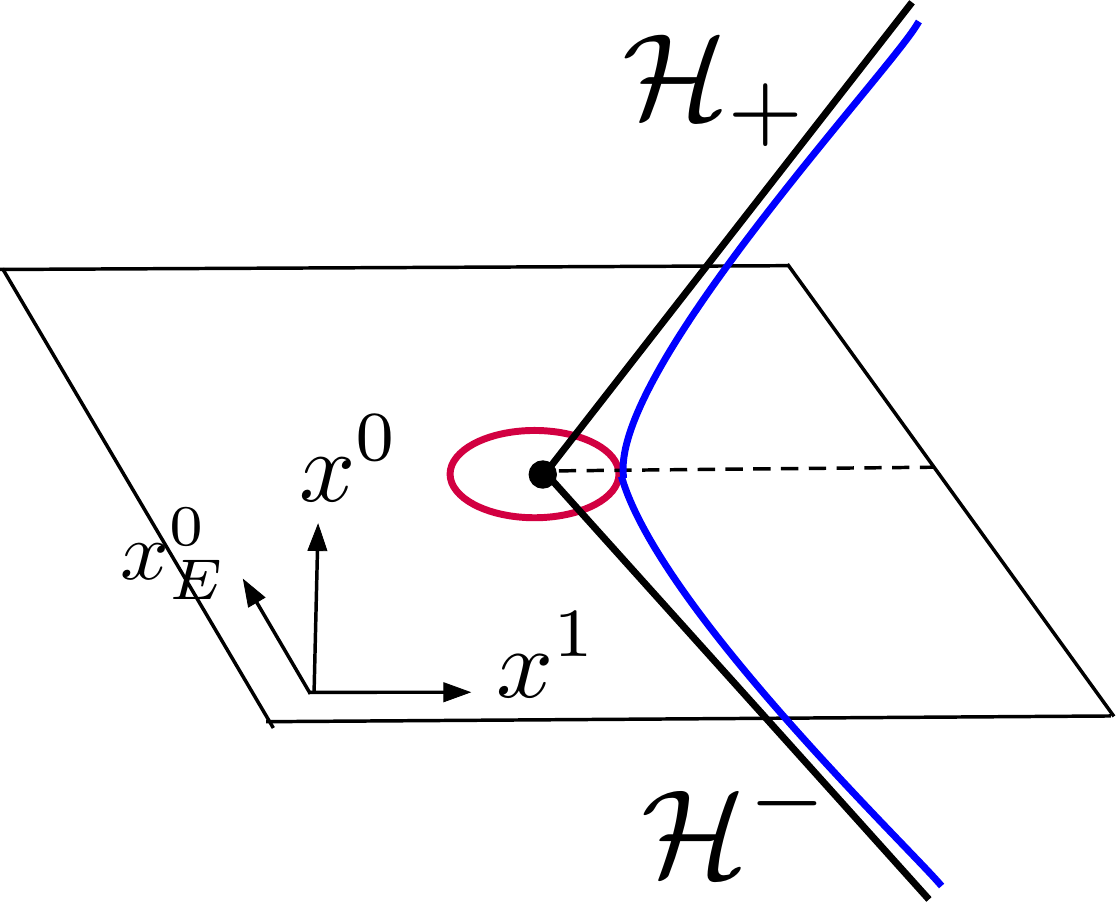} 
    \caption{\small{The plane denotes Euclidean space $(x^0_E, x^1)$, with the direction out of the plane denoting real time $x^0$ (transverse directions $\vec{x}$ are suppressed). In the Euclidean path integral, the shape deformation gives rise to a source involving the stress tensor integrated on an infinitesimal tube (red) of radius $a$ around the entanglement cut (black dot). The modular Hamiltonian is purely Lorentzian, and involves the stress tensor integrated on the Lorentzian surface shown in blue. In the limit where the cutoff $a$ goes to zero, the modular Hamiltonian can be expressed in terms of null energy operators intergrated on the future and past horizons $\mathcal{H}_{+}$ and $\mathcal{H}_-$ respectively. }}
    \label{fig:shape2}
\end{figure}

If  the original (undeformed) entanglement cut is at $(x^0, x^1)=(0,0)$, then we take the new deformed entanglement cut to lie at $(x^+,x^-) = (V^+(\vec{x}), V^-(\vec{x}))$, where $x^{\pm} = x^0 \pm x^1$ and $\vec{x}$ are coordinates along the original cut (see figure \ref{fig:shape}). The  point here is that we can always map the deformed subregion to the undeformed half-space inside the path-integral for the reduced density matrix by performing a local diffeomorphism in a small neighbourhood around the entanglement cut. Having done this, our subregion is now the original half-space, but the price we pay is that background metric gets deformed in this neighbourhood of the entanglement cut. Thus the action changes as:\footnote{There is an additional term of the form $\delta^{(2)}g_{\mu\nu}\sim \partial_{\mu}V^+\partial_{\nu}V^-$ in the metric deformation, which we will not keep track off. We expect this term to not contribute in the $a\to 0$ limit.}
\beq\label{ShapeDef}
\delta S = -\frac{1}{2}\int d^{d}x\,\delta g_{\mu\nu}(x)T^{\mu\nu}(x),\;\;\delta g_{\mu\nu}(x)= 2\partial_{(\mu}V_{\nu)}(x).
\eeq
In this way, we can represent a shape-deformation in terms of a source for the stress tensor close to the entanglement cut, thus enabling us to use equation \eqref{ModHamFinal} in order to get the modular Hamiltonian. Note further that because the source is a pure diffeomorphism, we can integrate by parts in the equation \eqref{ShapeDef}. Using the fact that the stress tensor is conserved, we can then rewrite this as
\beq
\delta S = -a\oint d^{d-1}x\, n^{\mu}V^{\nu} T^{\mu\nu}(x),
\eeq
where the integral is now over a cut-off tube of radius $a$ (with normal vector $n^{\mu}$) surrounding the entanglement cut, and we wish to take the limit $a\to 0$ in the end of the calculation (see figure \ref{fig:shape2}).\footnote{There are additional boundary terms from the cut surrounding the half-space region $R$, but these terms drop out of correlation functions (see \cite{Faulkner:2016mzt}) and will not be considered here.} So, the modular Hamiltonian for shape-deformed half-spaces is given by equation \eqref{ModHamFinal} after the replacement (see \cite{Faulkner:2016mzt} for details)
\beq \label{replace}
O(s_i) \to a\Big(-e^{s_i}T_{++} + e^{-s_i}T_{+-}\Big)e^{s_i+i\tau_i}V^+(\vec{x}_i) +a\Big(e^{-s_i}T_{--} - e^{s_i}T_{+-}\Big)e^{-s_i-i\tau_i}V^-(\vec{x}_i),
\eeq
with the stress tensors being inserted in real time at $(s_i, Y_i)$, and $a$ being the cutoff (see figure \ref{fig:shape2}). 

We can simplify the result by performing the $\tau_i$ integrals inside $d\mu_n$. Let us consider the term proportional to $V^{j_1}(\vec{x}_1)\cdots V^{j_n}(\vec{x}_n)$, where $j_i = \pm$. In this case, the $\tau$ integrals take the form:
\beqn
I_{j_1,\cdots,j_n} &=& \oint d\tau_1\cdots \oint d\tau_n e^{ j_1(s_1+i\tau_1)} \cdots e^{ j_n(s_n+i\tau_n)} f_{(n)}(s_1+i\tau_1,\cdots,s_n+i\tau_n) \\
&=& (-i)^n\oint dw_1\cdots\oint dw_n  \frac{(w_1e^{s_1})^{j_1}\cdots (w_ne^{s_n})^{j_n} e^{s_1}\cdots e^{s_n}}{(w_1e^{s_1}-1)(w_2e^{s_2}-w_1e^{s_1})\cdots(w_n e^{s_n}-w_{n-1}e^{s_{n-1}})(w_ne^{s_n}-1)}.\nonumber
\eeqn
where in the first line the exponential factors in front of $f_{(n)}$ come from the coefficients outside the parentheses in \eqref{replace}, and in the second line we have defined $w_i = e^{i\tau_i}$. The contours of integration here are $|w_i|=1$. We can simplify this integral by using
\beq \label{iden}
\frac{1}{(w_2e^{s_2}-w_1e^{s_1})}=\sum_{m=0}^{\infty}\left\{\Theta(s_2-s_1)\frac{w_1^me^{ms_1}}{w_2^{m+1}e^{(m+1)s_2}}-\Theta(s_1-s_2)\frac{w_2^me^{ms_2}}{w_1^{m+1}e^{(m+1)s_1}}\right\}.
\eeq
To simplify notation, let us define 
\beq
\Theta_n(s)=\begin{cases}\Theta(s) & \cdots n\geq 0 \\ -\Theta(-s)& \cdots n< 0. \end{cases}
\eeq
With this notation, the identity \eqref{iden} becomes
\beq
\frac{1}{(w_2e^{s_2}-w_1e^{s_1})}=\sum_{m=-\infty}^{\infty}\Theta_m(s_2-s_1)\frac{w_1^me^{ms_1}}{w_2^{m+1}e^{(m+1)s_2}},
\eeq
and so the integrals we must perform are
\beqn
I_{j_1,\cdots,j_n} &=&(-i)^n\oint dw_1\cdots\oint dw_n \frac{(w_1e^{s_1})^{j_1}\cdots (w_ne^{s_n})^{j_n} e^{s_1}\cdots e^{s_n}}{(w_1e^{s_1}-1)(w_ne^{s_n}-1)}\sum_{m_1=-\infty}^{\infty}\cdots \sum_{m_{n-1}=-\infty}^{\infty}\\
&\times & \Theta_{m_1}(s_2-s_1)\frac{(w_1e^{s_1})^{m_1}}{(w_2e^{s_2})^{m_1+1}}\Theta_{m_2}(s_3-s_2)\frac{(w_2e^{s_2})^{m_2}}{(w_3e^{s_3})^{m_2+1}}\cdots \Theta_{m_{n-1}}(s_n-s_{n-1})\frac{(w_{n-1}e^{s_{n-1}})^{m_{n-1}}}{(w_ne^{s_n})^{m_{n-1}+1}}.\nonumber
\eeqn
We can do the $w_2,\cdots, w_{n-1}$ integrals by the residue theorem, and this sets
\beq
m_2= m_1- j_2,\;\;m_3 = m_1 - (j_2+j_3),\cdots, m_{n-1}= m_1- (j_2+j_3+\cdots+ j_{n-1}).
\eeq
Finally doing the $w_1$ and $w_n$ integrals again using the Residue theorem
\beq
\frac{1}{2\pi i}\oint dw \frac{w^n}{(w-e^{-s})}=\Theta_n(s)e^{-ns},
\eeq 
we get (after redefining $m$)
\beqn
I_{j_1,\cdots,j_n}% &=&(2\pi)^n \sum_{m_1=-\infty}^{\infty} \Theta_{m_1+j_1}(s_1) \Theta_{m_1}(s_2-s_1)\Theta_{m_1-j_2}(s_3-s_2)\Theta_{m_1-j_2-j_3}(s_4-s_3)\cdots\nonumber\\
%&\times &\Theta_{m_1-j_2-j_3-\cdots-j_{n-1}}(s_n-s_{n-1})\Theta_{j_2+\cdots+j_n-m_1-1}(s_n)\nonumber\\
&=&(2\pi)^n \sum_{m=-\infty}^{\infty} \Theta_{m}(s_1)\Theta_{m-j_1}(s_2-s_1) \Theta_{m-(j_1+j_2)}(s_3-s_2)\Theta_{m-(j_1+j_2+j_3)}(s_4-s_3)\cdots\nonumber\\
&\times &\Theta_{m-(j_1+j_2+\cdots +j_{n-1})}(s_n-s_{n-1})\Theta_{(j_1+j_2+\cdots+j_n)-m-1}(s_n).
\eeqn

Having done the $\tau$ integrals, the modular Hamiltonian becomes
\beq 
K_{\lambda}= c_V+K +\sum_{n=1}^{\infty}\frac{1}{n!}\delta^nK ,
\eeq
\beq
\delta^nK=\frac{n!(-i)^{n-1}}{(2\pi)^{n-1}}\int d\vec{\mu}_n \sum_{j_1,\cdots,j_n}\int_{-\infty}^{\infty}ds_1\cdots\int_{-\infty}^{\infty}ds_n\, a^nI_{j_1\cdots j_n}(s_1,\cdots,s_n)\prod_{i=1}^nV^{j_i}(\vec{x}_i)\Big(-e^{s_i}T_{j_i,+} + e^{-s_i}T_{j_i,-}\Big),
\eeq
Now the only way this expression can be non-zero in the $a\to 0$ limit is that we pick up enhancements from $s_i$-integrals \cite{Faulkner:2015csl, Faulkner:2016mzt}. The $e^{s_i}T_{j_i,+}$ terms can only be enhanced when $s_i \to +\infty$ (more precisely from the regime $s_i \sim -\ln a$), so for these we can take the range of integration for $s_i$ to be $(0,\infty)$. It is further convenient to redefine $x^+_i = a e^{s_i}$. Similarly, the $e^{-s_i}T_{j_i,-}$ terms can only be enhanced when $s_i \to -\infty$ (from the regime $s_i \sim +\ln a$), and so take the range of integration for the corresponding $s_i$ to be $(-\infty, 0)$ and redefine $x^-_i = -a e^{-s_i}$. Having done these manipulations, we can send $a\to 0$ and the modular Hamiltonian becomes
\beq\label{ModHamShape}
\delta^nK=-2\pi\, n!i^{n-1} \sum_{j_1,\cdots,j_n}\sum_{k_1,\cdots,k_n}\int_{\mathcal{H}_{k_1}}V^{j_1}\,T_{j_1,k_1}\cdots \int_{\mathcal{H}_{k_n}}V^{j_n}\,T_{j_n,k_n}\,\mathcal{I}^{k_1\cdots k_n}_{j_1\cdots j_n}(x_1^{k_1},\cdots, x_n^{k_n}),
\eeq
where $k_i = \pm$, the integrals above are over future and past null horizons $\mathcal{H}_{\pm}$:
\beq
\int_{\mathcal{H}_{\pm}} = \int d^{d-2}\vec{x}_i \int_0^{\pm \infty}dx_i^{\pm},\;\;\;\; T_{j_i,\pm}=T_{j_i,\pm}(x_i^\pm,x_i^\mp=0,\vec{x}_i).
\eeq
and it is easy to convince oneself that we can simply replace $s_i \to x^{k_i}_i$ inside $I_{j_1,\cdots, j_n}$: 
\beqn
\mathcal{I}^{k_1\cdots k_n}_{j_1\cdots j_n}(x_1^{k_1},\cdots, x_n^{k_n})&=& \sum_{m=-\infty}^{\infty} \Theta_{m}(x_1^{k_1})\Theta_{m-j_1}(x^{k_2}_2 - x_1^{k_1})\cdots\\
&\times &\Theta_{m-(j_1+\cdots +j_{n-1})}(x_n^{k_{n}}-x_{n-1}^{k_{n-1}})\Theta_{(j_1+\cdots+j_n)-m-1}(x_n^{k_{n}}).\nonumber
\eeqn
Note that although the sum in the above expression seems to be over all integers, in practice it always truncates to a finite range. To better understand this, let us consider one term in equation \eqref{ModHamShape} for a particular fixed choice of $(j_1,\cdots, j_n)$ and $(k_1,\cdots,k_n)$. We can assume without loss of generality that
\beq 
(k_1\cdots, k_n)= \overbrace{++\cdots+}^{q_1}\overbrace{--\cdots-}^{q_2}\overbrace{++\cdots+}^{q_3}\cdots \overbrace{--\cdots-}^{q_{2M}},
\eeq
where $q_1$ and $q_{2M}$ are non-negative while all the other $q$s are strictly positive. Consider, for instance, the case $q_1>0,q_{2M}>0$. To obtain a non-zero answer the step functions impose the following constraints on the range of $m$:
\beq
 m \geq 0,\;\;m<\sum_{i=1}^{q_1}j_i,\;\;m\geq \sum_{i=1}^{q_1+q_2}j_i,\;\;\cdots\;\;,m < \sum_{i=1}^{q_1+q_2+\cdots +q_{2M-1}}j_i,\;\;m \geq \sum_{i=1}^{n}j_i.
 \eeq
Therefore the range of $m$ is given by
\beq
\mathrm{max}\left(0, \sum_{i=1}^{q_1+q_2}j_i,\cdots,\sum_{i=1}^{n}j_i\right) \leq m < \mathrm{min}\left( \sum_{i=1}^{q_1}j_i,\cdots,\sum_{i=1}^{q_1+\cdots+q_{2M-1}}j_i\right).
\eeq
Similarly, one can convince oneself that in the other three cases (i.e., with $q_1=0, q_{2M}>0$, etc.), the sum over $m$ always truncates to a finite sum. 
%We now have to perform the integrations over $\tau_{1,2}$:
%\beqn
%I_{\pm,\pm}(\sigma_1,\sigma_2) &=& \int d\tau_1\int d\tau_2 e^{\pm i\tau_1} e^{\pm i \tau_2} f \nonumber\\
%&=& -\oint \frac{dw_1}{w_1}\oint \frac{dw_2}{w_2} w_1^{\pm}w_2^{\pm} \frac{w_1e^{\sigma_1}w_2e^{\sigma_2}}{(w_1e^{\sigma_1}-1)(w_2e^{\sigma_2}-1)(w_2e^{\sigma_2}-w_1e^{\sigma_1})}\nonumber\\
%&=& -\oint dw_1\oint dw_2\, w_1^{\pm}w_2^{\pm} \frac{1}{(w_1-e^{-\sigma_1})(w_2-e^{-\sigma_2})(w_2e^{\sigma_2}-w_1e^{\sigma_1})}
%\eeqn
%We can simplify this integral by using
%\beq
%\frac{1}{(w_2e^{\sigma_2}-w_1e^{\sigma_1})}=\sum_{m=0}^{\infty}\left\{\Theta(\sigma_2-\sigma_1)w_1^mw_2^{-m-1}e^{m\sigma_1}e^{-(m+1)\sigma_2}-\Theta(\sigma_1-\sigma_2)w_2^mw_1^{-m-1}e^{m\sigma_2}e^{-(m+1)\sigma_1}\right\}.
%\eeq
%The integrals then factorize. Now we can use the contour integral:
%\beq
%\frac{1}{2\pi i}\oint dw \frac{w^n}{(w-e^{-s})}=\begin{cases}\Theta(s)e^{-ns} & \cdots n\geq 0 \\ -\Theta(-s)e^{-ns} & \cdots n< 0 \end{cases}
%\eeq
%and therefore
%\beq
%I_{++} = (4\pi)^2e^{-\sigma_1-\sigma_2}\Theta(\sigma_1)\Theta(\sigma_2)\mathrm{Sgn}(\sigma_2-\sigma_1),
%\eeq
%\beq
%I_{+-} = (4\pi)^2e^{-\sigma_1+\sigma_2}\Theta(\sigma_1)\Theta(-\sigma_2),
%\eeq
%\beq
%I_{-+} =- (4\pi)^2e^{\sigma_1-\sigma_2}\Theta(\sigma_2)\Theta(-\sigma_1),
%\eeq
%\beq
%I_{--} = (4\pi)^2e^{\sigma_1+\sigma_2}\Theta(-\sigma_1)\Theta(-\sigma_2)\mathrm{Sgn}(\sigma_2-\sigma_1),
%\eeq
%\subsection*{Second order}

The general expression \eqref{ModHamShape} is perhaps very abstract, so it is helpful to see the first few terms in this expansion. The first order result was already worked out in \cite{Faulkner:2016mzt}:
\beq
\delta^1K = -2\pi \int d^{d-2}\vec{x}\int_0^{\infty} dx^+\,V^+(\vec{x})T_{++}(x^+,\vec{x})-2\pi \int d^{d-2}\vec{x}\int_0^{-\infty} dx^-\,V^-(\vec{x})T_{--}(x^-,\vec{x}),
\eeq
and involves half-sided null energy operators. In \cite{Faulkner:2016mzt}, it was shown that this result, together with the monotonicty of relative entropy implies the averaged null energy condition. At second order, we find
\newcommand{\vx}{\vec{x}}
\beqn
\delta^2K &=&4\pi i\int_{\vec{x}_1}V^+(\vx_1)\int_{\vec{x}_2}V^+(\vx_2)\int_0^{\infty} dx_1^+ \int_0^{\infty}dx^+_2\,\Theta(x_1^+-x_2^+)\left[T_{++}(x_1^+,\vec{x}_1),T_{++}(x_2^+,\vec{x}_2)\right]\nonumber\\
&-&4\pi i\int_{\vec{x}_1}V^-(\vx_1)\int_{\vec{x}_2}V^-(\vx_2)\int_0^{-\infty} dx_1^- \int_0^{-\infty}dx^-_2\,\Theta(x_1^--x_2^-)\left[T_{--}(x_1^-,\vec{x}_1),T_{--}(x_2^-,\vec{x}_2)\right]\nonumber\\
&+& 4\pi i\int_{\vec{x}_1}V^+(\vx_1)\int_{\vec{x}_2}V^-(\vx_2)\int_0^{\infty} dx_1^+ \int_0^{-\infty}dx^-_2\,\left[T_{++}(x_1^+,\vec{x}_1),T_{--}(x_2^-,\vec{x}_2)\right].
\eeqn
Notice that the first two lines involve commutators of space-like separated operators, unless $\vec{x}_1= \vec{x}_2$. So we expect these to only contribute contact terms proportional to $\delta^{d-2}(\vec{x}_1- \vec{x}_2)$. As we will discuss below, this is a general feature of terms involving only $V^+$ or only $V^-$, i.e., purely null shape deformations. The last line on the other hand involves the commutator of the future half-sided null energy operator with the past half-sided null energy operator, and is not merely a contact term. It is easy enough to similarly work out higher order terms in the expansion.

\subsection{Null deformations}
The above formula simplifies greatly for the case of null-deformations, i.e., $V^-=0$ (or equivalently $V^+=0$). Consider for instance the second order result in this case:
\beq\label{null2}
\delta^2K =4\pi i\int_{\vec{x}_1}V^+(\vx_1)\int_{\vec{x}_2}V^+(\vx_2)\int_0^{\infty} dx^+_2 \int_{x_2^+}^{\infty}dx^+_1\,\left[T_{++}(x_1^+,\vec{x}_1),T_{++}(x_2^+,\vec{x}_2)\right].
\eeq
As we discussed above, the fact that the commutator is between space-like separated operators suggests that this is a pure contact term. We will now attempt to extract this contact term. We begin by writing
\beq \label{lim1}
 \int_{x_2^+}^{\infty}dx^+_1\,\left[T_{++}(x_1^+,\vec{x}_1),T_{++}(x_2^+,\vec{x}_2)\right] = A(x_2^+,\vec{x}_2) \delta(\vec{x}_1-\vec{x}_2).
\eeq
To deduce $A$, we can integrate along $\vec{x}_1$. However, we also need to regulate this expression. We will do so by pushing $T_{++}(x_2^+,\vec{x}_2)$ infinitesimally in the $-x_2^-$ direction (we can push it in any direction as long as we keep the operator inside the algebra of the original Rindler wedge):
\beqn \label{lim2}
A(x_2^+,\vec{x}_2)&=& \lim_{\epsilon \to 0}\int d^{d-2}\vec{x}_1\int_{x_2^+}^{\infty}dx^+_1\,\left[T_{++}(x_1^+,\vec{x}_1),T_{++}(x_2^+,x_2^-=-\epsilon,\vec{x}_2)\right] \nonumber\\
&=&\lim_{\epsilon \to 0}\int d^{d-2}\vec{x}_1\int_{-\infty}^{\infty}dx^+_1\,\left[T_{++}(x_1^+,\vec{x}_1),T_{++}(x_2^+,x_2^-=-\epsilon,\vec{x}_2)\right]\nonumber\\
&=& i\partial_+T_{++}(x_2^+,0,\vec{x}_2),
\eeqn
where in the second line the extra region in $x_1^+$ we added does not contribute because of space-like commutativity. So from equation \eqref{null2}, we get
\beqn
\delta^2K &=&- 4\pi \int_{\vec{x}_2}\Big(V^+(\vx_2)\Big)^2\int_0^{\infty} dx^+_2 \partial_+T_{++}(x_2^+,0,\vec{x}_2) \nonumber\\
&=& 2!(2\pi) \int_{\vec{x}_2}\Big(V^+(\vx_2)\Big)^2\,T_{++}(0,0,\vec{x}_2),
\eeqn
 where we have dropped the boundary term at $x^+_2\to\infty$.

In the general $n$th order case, we get from equation \eqref{ModHamShape}:
\beqn
\delta^nK&=&-2\pi\,n!i^{n-1}\prod_{i=1}^n \int_{\vec{x}_i}V^{+}(\vec{x}_i)\int_0^{\infty}dx_i^+ T_{++}(x^+_i,\vec{x}_i)\; \mathcal{I}^{++\cdots +}_{++\cdots +}(x_1^{+},\cdots, x_n^{+}),
\eeqn
where 
\beqn
\mathcal{I}^{++\cdots +}_{++\cdots+} &=& (-1)^{n-1}\Big\{\Theta(x_1^+-x_2^+)\Theta(x_2^+-x_3^+)\Theta(x_3^+-x_4^+)\cdots\Theta(x_{n-1}^+-x_n^+)\nonumber\\
&-&\Theta(x_2^+-x_1^+)\Theta(x_2^+-x_3^+)\Theta(x_3^+-x_4^+)\cdots\Theta(x_{n-1}^+-x_n^+)\nonumber\\
&+& \Theta(x_2^+-x_1^+)\Theta(x_3^+-x_2^+)\Theta(x_3^+-x_4^+)\cdots\Theta(x_{n-1}^+-x_n^+)\nonumber\\
&\vdots &  \nonumber\\
&+&(-1)^{n-1}\Theta(x_2^+-x_1^+)\Theta(x_3^+-x_2^+)\Theta(x_4^+-x_3^+)\cdots\Theta(x_n^+-x_{n-1}^+)\Big\}.
\eeqn
Note that all the integrals are on the future horizon; terms which involve integration on the past horizon vanish because the corresponding $\mathcal{I}$ function vanishes. Surprisingly, the structure of $\mathcal{I}^{++\cdots +}_{++\cdots +}$ precisely allows us to rewrite the above expression in terms of nested commutators:\footnote{We have checked this up to $n=9$ with a computer program. We also expect it to be provable using elementary techniques explained in appendix \ref{app:identity}, in particular  the discussion surrounding (\ref{eq:dag}). }
\beqn
\delta^nK&=&-2\pi\,n!(-i)^{n-1} \int_{\vec{x}_1}V^{+}(\vec{x}_1)\int_0^{\infty}dx_1^+\cdots \int_{\vec{x}_n}V^{+}(\vec{x}_n)\int_0^{\infty}dx_n^+\,\Theta(x_1^+-x_2^+)\Theta(x_2^+-x_3^+)\nonumber\\
&\cdots &\Theta(x_{n-1}^+-x_n^+)\left[\cdots\left[\left[T_{++}(x_1^+,\vec{x}_1),T_{++}(x_2^+,\vec{x}_2)\right],T_{++}(x_3^+,\vec{x}_3)\right]\cdots, T_{++}(x_n^+,\vec{x}_n)\right].
\eeqn
Finally, making repeated use of equations \eqref{lim1} and \eqref{lim2}, we obtain
\beq
\delta^n K = n!(2\pi) \int_{\vec{x}_n}\Big(V^+(\vx_n)\Big)^n\,\partial_+^{n-1}T_{++}(0,0,\vec{x}_n).
\eeq

In the present case, we can re-sum the perturbation series for the modular Hamiltonian into:
\beq\label{NullFinal}
K = 2\pi\int d^{d-2}\vec{x}\int_{V^+(\vec{x})}^{\infty}dx^+\,\left(x^+-V^+(\vec{x})\right)T_{++}(x^+,0,\vec{x}) + c_V. 
\eeq
This is our final expression for the modular Hamiltonian corresponding to null entanglement cuts. Equation \eqref{NullFinal} was conjectured in \cite{Faulkner:2016mzt}, and several arguments for the validity of this formula have appeared in \cite{Wall:2011hj, Bousso:2014uxa, Balakrishnan:2017bjg, Casini:2017roe, Lashkari:2017rcl, Koeller:2017njr}, based on free field theory, holography, and the algebraic theory of modular inclusions. Here, we have given another derivation of equation \eqref{NullFinal}, starting from our general result \eqref{ModHamShape} for shape-deformed modular Hamiltonians. We would like to mention some caveats though: firstly, we simply assumed that the commutator in equation \eqref{lim1} vanishes when $\vec{x}_1\neq \vec{x}_2$ because the operators are spacelike separated. However, this ignores potential subtleties at infinity, i.e. $x_{1,2}^+ \to \infty$ \cite{Kologlu:2019mfz}. Along the same lines, we dropped terms at infinity while integrating by parts in $x^+_i$. Finally, the regularization we used in equation \eqref{lim2} above seems reasonable to us, but it would be satisfying to give a more careful derivation by backing-up from the $a\to 0$ limit. We leave a more careful treatment of these issues to future work.

\section{Modular Flow near the Entanglement Cut}\label{sec:sec4}
In this section, we will use the formula we derived for the half-space modular Hamiltonian for Euclidean path integral states to study universal aspects of modular flow of local operators in conformal field theories (CFTs). One feature we will particularly be interested in is how the modular flow acts like a local boost in the limit the operator being flowed approaches the entanglement cut. We can use (vacuum) correlation functions of modular flowed operators as a probe of this physics. Since we want to extract universal information without specifying a lot of details about the CFT, we will only work to leading order in the state deformation. Let $\phi(Y)$ be a local operator with $Y=(x^1>0,\vec{x})$ being a point inside the half space, and define the modular flowed operator 
\beq
\phi_{\lambda}(s,Y) = e^{\frac{is}{2\pi}K_{\lambda}}\phi(0,Y)e^{\frac{is}{2\pi}K_{\lambda}}.
\eeq
Expanding in $\lambda$ to leading order, we find
\beq\label{Gf1}
\phi_{\lambda}(s,Y) = \phi(s,Y)-\frac{i}{2\pi}\int d\tau' dY' \lambda(\tau',Y')\int_{-\infty}^{\infty}ds'g(s,s'+i\tau')\left[O(s',Y'), \phi(s,Y)\right]+O(\lambda^2),
\eeq
where $\phi(s,Y) = e^{\frac{is}{2\pi}K}\phi(0,Y)e^{\frac{is}{2\pi}K}$ is the operator flowed with the vacuum modular Hamiltonian, and we have denoted
\beq
g(s,s')=\frac{f_{(2)}(s',s)}{f_{(1)}(s)} =\frac{1}{2}\frac{\sinh(s/2)}{\sinh(s'/2)\sinh((s-s')/2)}.
\eeq
%For simplicity, we consider a state deformation obtained from turning on a source for the same operator $O$ in the Euclidean path integral. Performing the $t$ integration in equation \eqref{ML}, we land at
%\beq
%\partial_{\lambda}G = i\int d^dz\,\lambda(z)\int_{-\infty}^{\infty}ds'\,g(s,s'+i\tau_z)\langle 0|\left[\left[O^{(s')}(\tilde{z}),O^{(s)}(x)\right],O(y)\right]|0\rangle,
%\eeq
%where if $z=(z_1\sin\tau_z, z_1\cos\tau_z,\vec{z})$ then $\tilde{z}=(0,z_1,\vec{z})$, and
%\beq
%g(s,s')=\frac{1}{2}\frac{\sinh(s/2)}{\sinh(s'/2)\sinh((s'-s)/2)}.
%\eeq
We can probe the structure of $\phi_{\lambda}$ by computing its correlation functions with other local operators in the undeformed vacuum:
\beq
G_{\lambda}(s)=\langle 0 | \phi_{\lambda}(s,x^1,\vec{x})\phi(0,y^1,\vec{y})|0\rangle.
\eeq
From equation \eqref{Gf1} we then obtain
\beq\label{Gf2}
G_{\lambda} = G_0-\frac{i}{2\pi}\int d\mu_1\int_{-\infty}^{\infty}ds'\,g(s,s'+i\tau')\langle 0|\left[O(s',z^1,\vec{z}),\phi(s,x^1,\vec{x})\right]\phi(0,y^1,\vec{y})|0\rangle+O(\lambda^2),
\eeq
where
$$d\mu_1= d\tau' dz^1d^{d-2}\vec{z}\; \lambda(\tau',z^1,\vec{z}).$$
\begin{figure}[t]
    \centering
    \includegraphics[height=5.2cm]{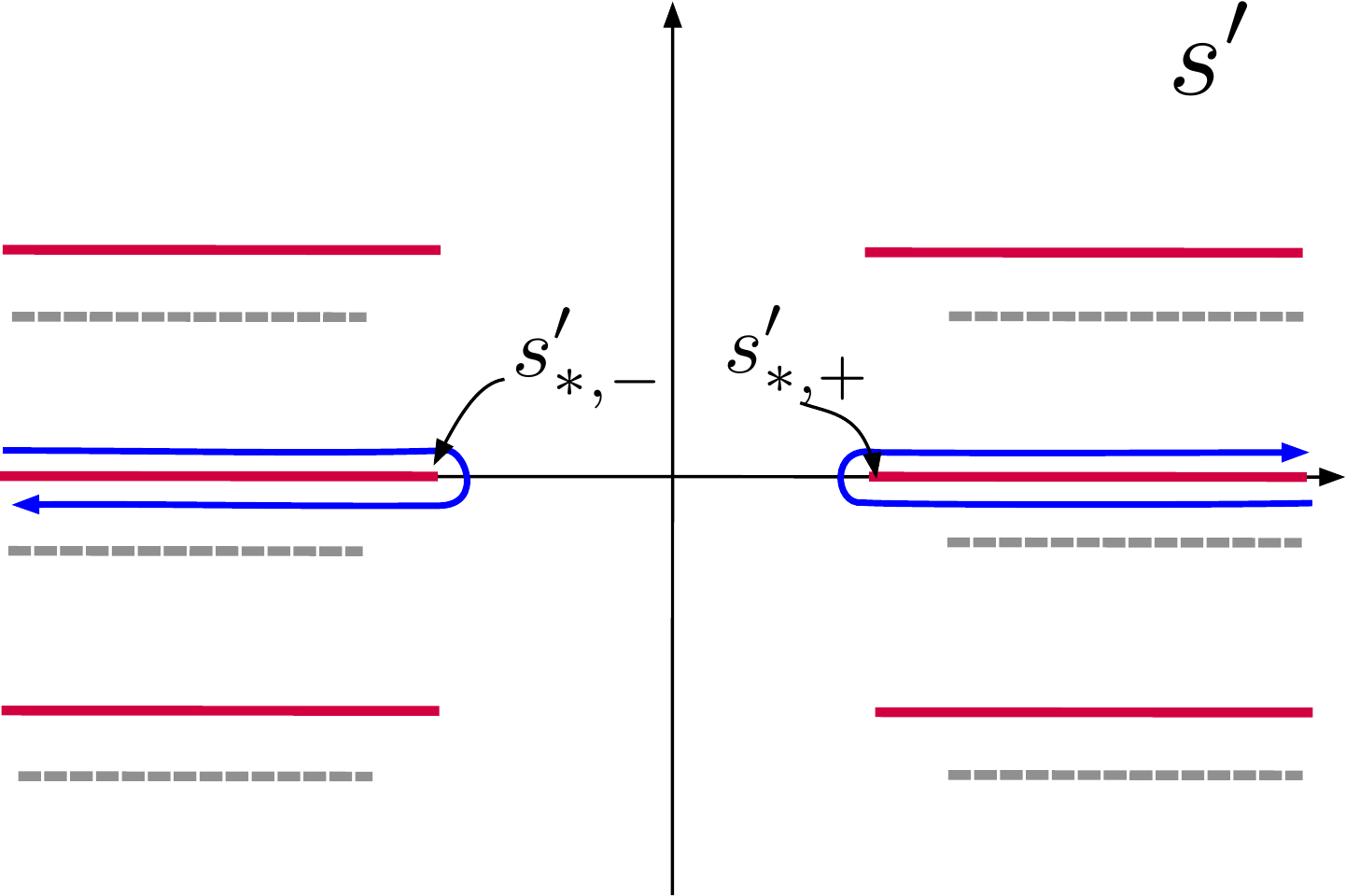} 
    \caption{\small{In the $s'$ complex plane, the correlation function has branch cuts due to the operator at $(s,x^1,\vec{x})$ (shown in red) and due to the operator at $(y^1,\vec{y})$ (shown in grey), which are displaced by an infinitesimal amount below the red branch cuts (exaggerated in the figure). These are repeated with periodicity $2\pi$ along the imaginary axis. The contour $\Gamma$ for $s'$-integral is shown in blue. The poles due to $g$ are not shown here. }}
    \label{fig:contour}
\end{figure}

\begin{figure}
    \centering
 \includegraphics[height=5.2cm]{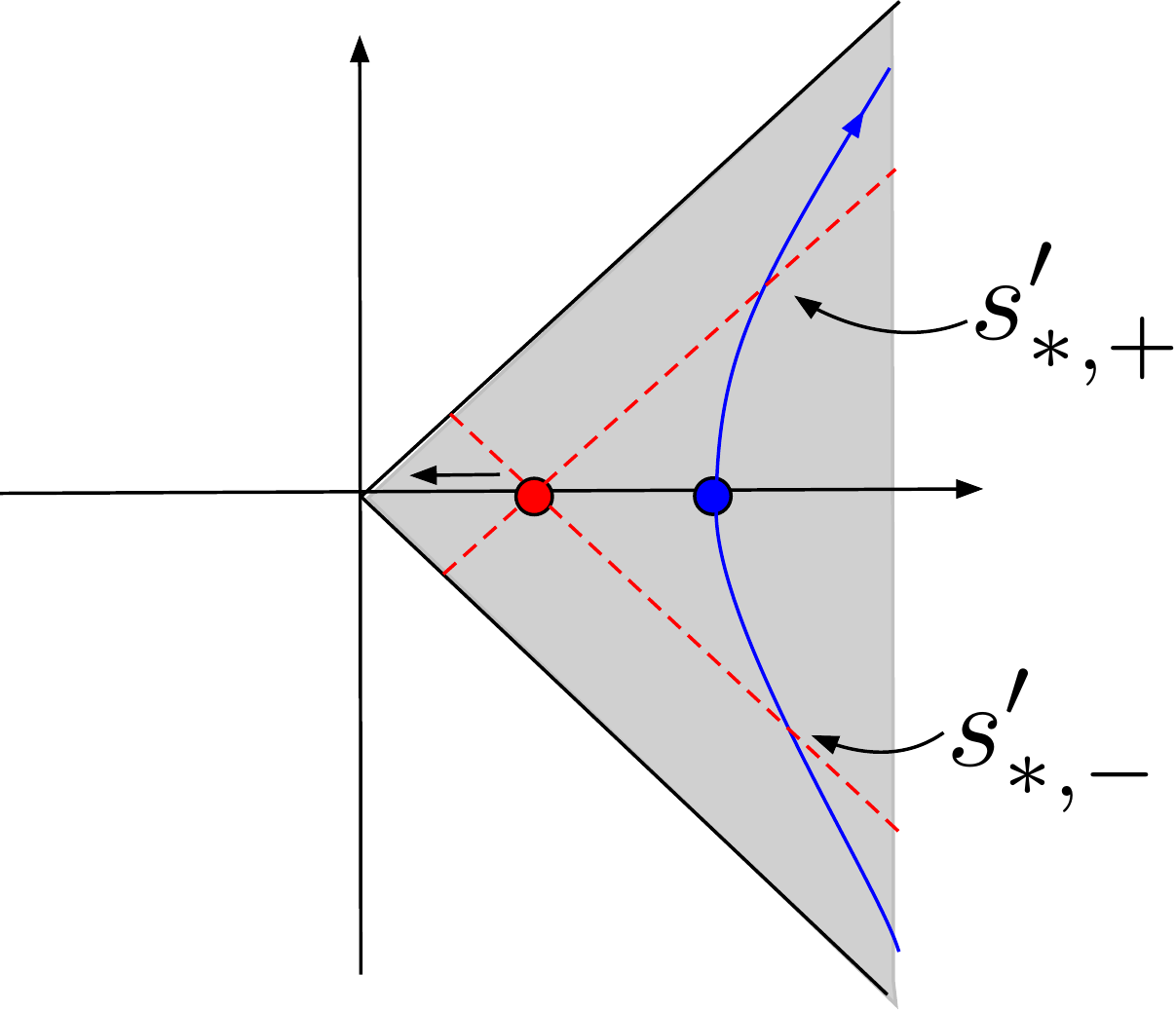}
    \caption{\small{As the operator being flowed (shown in red) approaches the entanglement cut, the contribution to the $s'$ integral comes from larger and larger $|s'|$. This happens because the commutator with the sourced operator (blue) vanishes outside the light-cone of the flowed operator (dashed red lines).}}
    \label{fig:flow}
\end{figure}

For simplicity, let us focus on the case where $\phi = O$. The Euclidean three point function in any CFT is universal and given by
\beq \label{3pt}
\langle O(x)O(y)O(z) \rangle =\frac{D_{\Delta}}{|x-y|^{\Delta}|y-z|^{\Delta}|z-x|^{\Delta}},
\eeq
where $\Delta$ is the conformal dimension of $O$ and $D_{\Delta}$ is a constant. We can obtain the Lorentzian three point function in \eqref{Gf2} from here by analytic continuation. In particular, the commutator can be obtained from a branch discontinuity of the three point function (see figure \ref{fig:contour}): %For instance, we can compute the commutator of interest as
%\beq\label{comm1}
%\langle \left[O(\tilde{z},s'),O(x,s)\right]O(y) \rangle =f_{\Delta}(x_s,y,z_{s'})\left\{d_+\Theta(s'-s'_{*,+})+d_-\Theta(s'_{*,-}-s')\right\}.
%\eeq
%where $f_{\Delta}$ is the same function appearing on the RHS of equation \eqref{3pt}, $d_{\pm}$ are the branch discontinuity factors across the cuts, and $s_*^{\pm}$ are the locations of the lightcone singularities when the two operators become light-like separated:
%\beq
%s'_{*,\pm} = s+ \log\left(\alpha \pm \sqrt{\alpha^2-1}\right),\;\;\;\alpha = \frac{(\vec{z}-\vec{x})^2+x_1^2+z_1^2}{2x_1 z_1}.
%\eeq
\beq
\delta G = \frac{-iD_{\Delta}}{2\pi[(\vec{x}-\vec{y})^2+ x_1^2+y_1^2-2x_1y_1\cosh(s+i\epsilon)]^{\Delta/2}}I
\eeq
\beq
I = \int_{\Gamma}ds'\frac{g(s,s'+i\tau')}{\left[(\vec{x}-\vec{z})^2+x_1^2+z_1^2-2x_1z_1\cosh(s'-s)\right]^{\Delta/2}\left[(\vec{y}-\vec{z})^2+y_1^2+z_1^2-2y_1z_1\cosh(s'+i\epsilon)\right]^{\Delta/2}},
\eeq
where the contour $\Gamma$ is shown in figure \ref{fig:contour} and surrounds the branch-cuts due to the operator at $(s,x^1,\vec{x})$. These cuts begin at the locations of the lightcone singularities when the two operators become light-like separated (see figure \ref{fig:flow}):
\beq
s'_{*,\pm} = s+ \log\left(\alpha \pm \sqrt{\alpha^2-1}\right),\;\;\;\alpha = \frac{(\vec{z}-\vec{x})^2+x_1^2+z_1^2}{2x_1 z_1},
\eeq
and are repeated with period $2\pi$ along the imaginary axis. Note that the contour $\Gamma$ has two pieces, which we may denote $\Gamma_{\pm}$ coming from the future and past cuts respectively. 

Let us now consider what happens when the operator being flowed starts approaching the entanglement cut, i.e., $x^1 \to 0$ (with everything else fixed). The calculation simplifies in this limit as in this case, $\alpha \to \infty$, and correspondingly $s_{*,\pm}' \to \pm \log (2\alpha).$ The key physical point is that as $x^1\to 0$, the commutator in equation \eqref{Gf1} or \eqref{Gf2} only receives contribution for larger and larger $|s'|$, $s'\in (-\infty, s'_{*,-}] \cup [s'_{*,+},\infty)$ to be specific; it vanishes outside this domain because the operators are spacelike separated (see figure \ref{fig:flow}). The function $g$ decays exponentially in $s'$ in this regime:
\beq
g\sim -2 e^{\pm s/2}\sinh(s/2)e^{\mp (s'+i\tau')}.
\eeq
Furthermore, the correlation function also decays exponentially in $s'$ in this regime, with the end result that the integral is suppressed. To see this a bit more explicitly, consider the $\Gamma_+$ contour. We define
$$\Lambda = e^{s'-s_{*,+}'}.$$
The leading order (in $x_1$) contribution to the integral becomes
\beq
I_+ = (x_1/y_1)^{\Delta/2+1}\frac{2y_1z_1}{\left[(\vec{x}-\vec{z})^2+z_1^2\right]^{\Delta+1}}e^{-\frac{\Delta+1}{2}s-i\tau_z}\sinh(s/2)\int_{\Gamma^{(\Lambda)}_+}\frac{d\Lambda}{\Lambda^2}\frac{e^{iQ}}{\left|\Lambda(\Lambda-1)\right|^{\Delta/2}}+\cdots,
%&=&(x_1/y_1)^{\Delta/2+1}\frac{2y_1z_1}{\left[(\vec{x}-\vec{z})^2+z_1^2\right]^{\Delta+1}}e^{-\frac{\Delta+1}{2}s-i\tau_z}\sinh(s/2)\frac{\Gamma(\Delta+1)\Gamma(1-\Delta/2)}{\Gamma(\Delta/2+2)}+\cdots.
\eeq
where $\Gamma^{(\Lambda)}_+$ is the new contour for $\Lambda$ which straddles the branch cut from $(1,\infty)$, and $Q$ is a pure phase (which is different above and below the cut). Clearly this contribution is suppressed as $x_1\to 0$. Similarly the contribution from $\Gamma_-$ is also suppressed. Therefore, the leading order correction to the modular flow vanishes as $x^1 \to 0$, i.e., as the operator being flowed approaches the entanglement cut. Thus we get the result that  modular flow using the perturbed modular hamiltonian is identitcal to the vacuum  half-space modular flow in this limit. Although we have shown this only to leading order in $\lambda$, we expect this result to be true to all orders because at higher orders as well, since modular flow should involve commutators between the sourced operators and the flowed operator, and as the flowed operator approaches the entanglement cut a similar suppression of the commutator should occur. We leave these details to future work. 

\section{Discussion}
In this paper, we derived a manifestly Lorentzian formula for the Rindler modular Hamiltonian for a class of excited states constructed by turning on sources in the Euclidean path integral. When applied to the case of shaped deformed subregions, the formula gives an explicit expression for the (vaccum) modular Hamiltonian to all orders in the shape deformation. In the special case of null deformations, the series can be resummed and gives precisely the known formula for the modular Hamiltonian for null cuts of the Rindler horizon. We also used our results at leading order to demonstrate how modular flow for even excited states acts like a local boost near the entanglement cut.

Euclidean path integral states are of direct relevance in the AdS/CFT correspondence \cite{Marolf:2017kvq}. It would be interesting to see whether our all-orders formula for the modular Hamiltonian can be translated in the language of bulk gravitational quantities in holographic CFTs. If so, this could give a purely \emph{Lorentzian} derivation of the Jafferis-Lewkowycz-Maldacena-Suh (JLMS) formula, and thus indeed the Ryu-Takayanagi formula, at least up to $O(G_N^0)$. We hope that this could also shed light on subleading (in $G_N$) corrections to the JLMS formula \cite{Dong:2017xht}. Another potential application to AdS/CFT is in deriving the the non-linear gravitational equations of motion in the bulk from the physics of entanglement in the CFT. The leading order version of our formula played a crucial role in obtaining the second order gravitational equations of motion in \cite{Faulkner:2017tkh, Haehl:2017sot}. It seems natural that the fully non-linear gravitational equations are hidden in our all orders formula for the modular Hamiltonian. It would also be interesting to see whether our analysis can be extended to include ``multi-trace'' operators (i.e., multi-local  operators of the form $O(x_1)\cdots O(x_n)$) in the Euclidean path-integral. At face value, it appears that the perturbation theory in this case is ill-defined (see \cite{Haehl:2019fjz} for a detailed discussion) and inevitably leads to out-of-Euclidean-time-ordered correlators. However, in conformal field theories, it may be possible to first use the operator product expansion (OPE) to rewrite the multi-trace operator in terms of local operators and then use our results to obtain the modular Hamiltonian. This is not always guaranteed to work (i.e., the OPE may not converge), but it does seem to be a useful way forward in some cases of interest.

Our all-orders formula for the shape deformed modular Hamiltonian involved various products of (half-sided) integrated null energy operators. These are reminiscent of the light-ray operators which have been studied recently in \cite{Kravchuk:2018htv, Cordova:2018ygx, Kologlu:2019mfz, Balakrishnan:2019gxl, Belin:2019mnx} from a bootstrap perspective. It is plausible that the technology of light ray operators can lead to further progress in understanding properties of modular flow. Relatedly, we hope that the formula for the modular Hamiltonian for excited states derived here can shed some light on operator growth under modular flow, modular chaos \cite{Faulkner:2018faa, deBoer:2019uem} and similar questions. Finally, it would be interesting to find situations where our perturbative formula for the modular Hamiltonian can be re-summed. The vacuum modular Hamiltonian for null cuts of the Rindler horizon provides one such example, but it may be possible to engineer non-trivial excited states where such a re-summation is possible.

    \section*{Acknowledgements}
We are grateful to Tom Faulkner for several helpful conversations. We would also like to thank Manthos Karydas, G\'abor S\'arosi and Tomonori Ugajin for discussions and helpful comments on the manuscript.

\appendix 
\section{Proof of combinatorial identity (\ref{eq:cancel}) }
\label{app:identity}
We begin with describing the closed form result for the quantity $c(p)$. Let $p \equiv (p_1,p_2, \ldots , p_k)$ be an ordered partition of $b+1$ with $p_1 + p_2 + \ldots + p_k = b+1$. Then 
\beq
c(p)  = (-1)^{k+1} p_k  \frac{b!}{\mathbf{n}(p)}.
\label{eq:cans}
\eeq
The above expression is clearly true for $b=0$ (the base case). We now verify that it also satisfies the recursion (\ref{eq:crecursion}). Inductively assume that an analogous formula is true for all ordered partitions of $b$. Then the RHS of (\ref{eq:crecursion}) can be computed. We have
\beq
c(p) = (-1)^{k+1}(b-1)! p_k \left(  \sum_{i=1}^{k-1} \frac{p_i}{\mathbf{n}(p)}\right)   +  (-1)^{k+1}(b-1)! (p_k -1) \frac{p_k}{\mathbf{n}(p) }.
\eeq
To be more precise, we infact need to consider two separate cases $p_k>1$ and $p_k=1$, but in both cases the above expression holds due to the details of the sets $\mathcal{P}_+(p)$ and $\mathcal{P}_-(p)$ explained in the paragraph below (\ref{eq:crecursion}).
The RHS above simply evaluates to  (\ref{eq:cans}) when rewritten using $\sum_{i=1}^{k} p_i = b+1$ which is all that we need to check to inductively prove (\ref{eq:cans}).

We refer to $p_k$ in the computation above as the ``last part''. We call the first term in (\ref{eq:cancel}), the $\tilde{P}$-term and the second term, the $T$-term. Plugging in the formula for $c(p)$, we see that the summation in $T$-term can be organized according to last part as follows
\begin{align}
T\text{-term} &\equiv \sum_{p \in \mathcal{P}_{b+1} } \frac{c(p) \mathbf{n}(p) }{b!} \mathbf{\Theta}_p(\tau_0,\tau_1,\ldots , \tau_b) \\
&= \sum_{j=1}^{b+1} j (-1)^j  \left( \prod_{i=1}^{b-j} \left( 1- \Theta(\tau_{i-1} - \tau_{i} ) \right) \right) \Theta( (\tau_{b-j+1}, \tau_{b-j+2}, \ldots b ) ) \\
&= \sum_{j=1}^{b+1} j(-1)^j \Theta((\tau_{b-j}, \tau_{b-j-1} , \ldots \tau_1) ) \Theta((\tau_{b-j+1},\tau_{b-j+2},\ldots, \tau_b))
\label{eq:Tterm1}
\end{align}
In the second line above, $j$ labels the values taken by the last part which range from $1$ to $b+1$.  In the third line we have used the identity $\Theta(\tau_i - \tau_j) +\Theta(\tau_j-\tau_i) = 1$.

We now use the following general trick involving step functions. A nonzero product of step function of $(b+1)$~ $\tau$-variables can be represented as a directed acyclic graph(DAG) where the vertices are the indices on the $\tau$-variables and there is an edge from $i$ to $j$ for each step function $\Theta(\tau_i - \tau_j)$ in the product. Such a DAG representing a product of $\Theta$'s can be replaced with a sum over all its ``topological sortings'' (permutations of all the indices such that edges run strictly from left to right). Applying this to the above expression, we can write
\beq
\Theta((\tau_{b-j}, \tau_{b-j-1} , \ldots \tau_1) ) \Theta(\tau_{b-j+1},\tau_{b-j+2},\ldots, \tau_b) = \sum_q  \Theta((\tau_{q_0},\tau_{q_1},\ldots \tau_{q_b} )),
\eeq
where the permutations $q$ appearing in the above sum are all possible topological sortings of the following directed graph:
\begin{align}
\boxed{b-j+1}\rightarrow \boxed{b-j+2}\rightarrow \ldots \rightarrow \boxed{b} \nonumber \\
\boxed{b-j}\rightarrow \boxed{b-j-1}\rightarrow \ldots \rightarrow \boxed{1} 
\label{eq:dag}
\end{align}
We can now make a one-to-one correspondence between all topological sortings of above graph and strings $\alpha \in \{ \rni, \rnii \}^{b+1} $ which contain $j$ occurences of $\rni$. Scan the string $\alpha$ from left to right. If the current letter is $\rni$, pick the leftmost unpicked vertex from the first line of the graph (\ref{eq:dag}), and if the current letter $\rnii$, then we pick the left most unpicked vertex from the second line of the graph. Clearly the set of permutations obtained this way is identical to the set of permutations in the $\tilde{P}$-term and we have chosen the labels $\rni,\rnii$ in a analogous fashion here (see (\ref{eq:taupermutation})). There is a slight caveat, which is that as written here, there are  $2^{b+1}$ strings to deal with in $T$-term, while there are $2^{b}$ strings  in $\tilde{P}$-term. This is dealt with as follows: Denote by  $q_\alpha$ the permutation obtained in this process for a string $\alpha \in \{\rni,\rnii\}^{b+1}$, and by $r_\alpha$ the number of $\rni$'s in $\alpha$. For every string $\alpha \in \{\rni,\rnii\}^{b+1}$  such that $\alpha = (\rni  \beta)$ where $\beta\in \{ \rni, \rnii \}^{b}$, the string $\alpha' = (\rnii \beta) \in \{ \rni,\rnii\}^{b+1}$ gives rise to the same permutation as $\alpha$, i.e., $q_\alpha = q_{\alpha'}$. Thus, for $\alpha = (\rni \beta)$, the total contribution multiplying $\Theta(\vec{\tau}(q_\alpha))$ is  given by $\left(r_\alpha (-1)^{r_\alpha} + (r_\alpha - 1) (-1)^{r_\alpha - 1}\right) = (-1)^{r_\alpha}$ (where we have included also the contribution from $(\rnii \beta)$).
These facts when plugged into (\ref{eq:Tterm1}) yields
\beq
T\text{-term} = \sum_{\alpha \in \left\{ \{\rni, \rnii \}^{b+1} \big\vert  \text{first letter}=\rni \right\} }  (-1)^{r_\alpha} \Theta( (\tau_{q_\alpha(0) } , \tau_{q_\alpha(1) }, \ldots, \tau_{q_\alpha(b) } )).
\eeq
Cancellation is now evident by pairing up $(b+1)$-length strings which begin with letter $\rni$ in the obvious fashion with the $b$-length strings that label the summation in $\tilde{P}$-term.

\providecommand{\href}[2]{#2}\begingroup\raggedright\endgroup


\providecommand{\href}[2]{#2}\begingroup\raggedright\begin{thebibliography}{10}

\bibitem{Faulkner:2016mzt}
T.~Faulkner, R.~G. Leigh, O.~Parrikar and H.~Wang, \emph{{Modular Hamiltonians
  for Deformed Half-Spaces and the Averaged Null Energy Condition}},
  \href{http://dx.doi.org/10.1007/JHEP09(2016)038}{\emph{JHEP} {\bf 09} (2016)
  038}, [\href{https://arxiv.org/abs/1605.08072}{{\tt 1605.08072}}].

\bibitem{Balakrishnan:2017bjg}
S.~Balakrishnan, T.~Faulkner, Z.~U. Khandker and H.~Wang, \emph{{A General
  Proof of the Quantum Null Energy Condition}},
  \href{http://dx.doi.org/10.1007/JHEP09(2019)020}{\emph{JHEP} {\bf 09} (2019)
  020}, [\href{https://arxiv.org/abs/1706.09432}{{\tt 1706.09432}}].

\bibitem{Ceyhan:2018zfg}
F.~Ceyhan and T.~Faulkner, \emph{{Recovering the QNEC from the ANEC}},
  \href{https://arxiv.org/abs/1812.04683}{{\tt 1812.04683}}.

\bibitem{Jafferis:2015del}
D.~L. Jafferis, A.~Lewkowycz, J.~Maldacena and S.~J. Suh, \emph{{Relative
  entropy equals bulk relative entropy}},
  \href{http://dx.doi.org/10.1007/JHEP06(2016)004}{\emph{JHEP} {\bf 06} (2016)
  004}, [\href{https://arxiv.org/abs/1512.06431}{{\tt 1512.06431}}].

\bibitem{Ryu:2006bv}
S.~Ryu and T.~Takayanagi, \emph{{Holographic derivation of entanglement entropy
  from AdS/CFT}},
  \href{http://dx.doi.org/10.1103/PhysRevLett.96.181602}{\emph{Phys. Rev.
  Lett.} {\bf 96} (2006) 181602},
  [\href{https://arxiv.org/abs/hep-th/0603001}{{\tt hep-th/0603001}}].

\bibitem{Hubeny:2007xt}
V.~E. Hubeny, M.~Rangamani and T.~Takayanagi, \emph{{A Covariant holographic
  entanglement entropy proposal}},
  \href{http://dx.doi.org/10.1088/1126-6708/2007/07/062}{\emph{JHEP} {\bf 07}
  (2007) 062}, [\href{https://arxiv.org/abs/0705.0016}{{\tt 0705.0016}}].

\bibitem{Faulkner:2013ana}
T.~Faulkner, A.~Lewkowycz and J.~Maldacena, \emph{{Quantum corrections to
  holographic entanglement entropy}},
  \href{http://dx.doi.org/10.1007/JHEP11(2013)074}{\emph{JHEP} {\bf 11} (2013)
  074}, [\href{https://arxiv.org/abs/1307.2892}{{\tt 1307.2892}}].

\bibitem{Engelhardt:2014gca}
N.~Engelhardt and A.~C. Wall, \emph{{Quantum Extremal Surfaces: Holographic
  Entanglement Entropy beyond the Classical Regime}},
  \href{http://dx.doi.org/10.1007/JHEP01(2015)073}{\emph{JHEP} {\bf 01} (2015)
  073}, [\href{https://arxiv.org/abs/1408.3203}{{\tt 1408.3203}}].

\bibitem{Dong:2016eik}
X.~Dong, D.~Harlow and A.~C. Wall, \emph{{Reconstruction of Bulk Operators
  within the Entanglement Wedge in Gauge-Gravity Duality}},
  \href{http://dx.doi.org/10.1103/PhysRevLett.117.021601}{\emph{Phys. Rev.
  Lett.} {\bf 117} (2016) 021601},
  [\href{https://arxiv.org/abs/1601.05416}{{\tt 1601.05416}}].

\bibitem{Harlow:2016vwg}
D.~Harlow, \emph{{The Ryu?Takayanagi Formula from Quantum Error Correction}},
  \href{http://dx.doi.org/10.1007/s00220-017-2904-z}{\emph{Commun. Math. Phys.}
  {\bf 354} (2017) 865--912}, [\href{https://arxiv.org/abs/1607.03901}{{\tt
  1607.03901}}].

\bibitem{Faulkner:2017vdd}
T.~Faulkner and A.~Lewkowycz, \emph{{Bulk locality from modular flow}},
  \href{http://dx.doi.org/10.1007/JHEP07(2017)151}{\emph{JHEP} {\bf 07} (2017)
  151}, [\href{https://arxiv.org/abs/1704.05464}{{\tt 1704.05464}}].

\bibitem{Faulkner:2018faa}
T.~Faulkner, M.~Li and H.~Wang, \emph{{A modular toolkit for bulk
  reconstruction}},
  \href{http://dx.doi.org/10.1007/JHEP04(2019)119}{\emph{JHEP} {\bf 04} (2019)
  119}, [\href{https://arxiv.org/abs/1806.10560}{{\tt 1806.10560}}].

\bibitem{Cotler:2017erl}
J.~Cotler, P.~Hayden, G.~Penington, G.~Salton, B.~Swingle and M.~Walter,
  \emph{{Entanglement Wedge Reconstruction via Universal Recovery Channels}},
  \href{http://dx.doi.org/10.1103/PhysRevX.9.031011}{\emph{Phys. Rev.} {\bf X9}
  (2019) 031011}, [\href{https://arxiv.org/abs/1704.05839}{{\tt 1704.05839}}].

\bibitem{Czech:2019vih}
B.~Czech, J.~De~Boer, D.~Ge and L.~Lamprou, \emph{{A modular sewing kit for
  entanglement wedges}},
  \href{http://dx.doi.org/10.1007/JHEP11(2019)094}{\emph{JHEP} {\bf 11} (2019)
  094}, [\href{https://arxiv.org/abs/1903.04493}{{\tt 1903.04493}}].

\bibitem{2006PhRvD..74f6009H}
A.~{Hamilton}, D.~{Kabat}, G.~{Lifschytz} and D.~A. {Lowe}, \emph{{Holographic
  representation of local bulk operators}},
  \href{http://dx.doi.org/10.1103/PhysRevD.74.066009}{\emph{Phys. Rev. D} {\bf
  74} (Sep, 2006) 066009}, [\href{https://arxiv.org/abs/hep-th/0606141}{{\tt
  hep-th/0606141}}].

\bibitem{Nozaki:2013vta}
M.~Nozaki, T.~Numasawa, A.~Prudenziati and T.~Takayanagi, \emph{{Dynamics of
  Entanglement Entropy from Einstein Equation}},
  \href{http://dx.doi.org/10.1103/PhysRevD.88.026012}{\emph{Phys. Rev.} {\bf
  D88} (2013) 026012}, [\href{https://arxiv.org/abs/1304.7100}{{\tt
  1304.7100}}].

\bibitem{Faulkner:2013ica}
T.~Faulkner, M.~Guica, T.~Hartman, R.~C. Myers and M.~Van~Raamsdonk,
  \emph{{Gravitation from Entanglement in Holographic CFTs}},
  \href{http://dx.doi.org/10.1007/JHEP03(2014)051}{\emph{JHEP} {\bf 03} (2014)
  051}, [\href{https://arxiv.org/abs/1312.7856}{{\tt 1312.7856}}].

\bibitem{Faulkner:2017tkh}
T.~Faulkner, F.~M. Haehl, E.~Hijano, O.~Parrikar, C.~Rabideau and
  M.~Van~Raamsdonk, \emph{{Nonlinear Gravity from Entanglement in Conformal
  Field Theories}},
  \href{http://dx.doi.org/10.1007/JHEP08(2017)057}{\emph{JHEP} {\bf 08} (2017)
  057}, [\href{https://arxiv.org/abs/1705.03026}{{\tt 1705.03026}}].

\bibitem{Haehl:2017sot}
F.~M. Haehl, E.~Hijano, O.~Parrikar and C.~Rabideau, \emph{{Higher Curvature
  Gravity from Entanglement in Conformal Field Theories}},
  \href{http://dx.doi.org/10.1103/PhysRevLett.120.201602}{\emph{Phys. Rev.
  Lett.} {\bf 120} (2018) 201602},
  [\href{https://arxiv.org/abs/1712.06620}{{\tt 1712.06620}}].

\bibitem{Lewkowycz:2018sgn}
A.~Lewkowycz and O.~Parrikar, \emph{{The holographic shape of entanglement and
  Einstein’s equations}},
  \href{http://dx.doi.org/10.1007/JHEP05(2018)147}{\emph{JHEP} {\bf 05} (2018)
  147}, [\href{https://arxiv.org/abs/1802.10103}{{\tt 1802.10103}}].

\bibitem{Bisognano:1976za}
J.~J. Bisognano and E.~H. Wichmann, \emph{{On the Duality Condition for Quantum
  Fields}}, \href{http://dx.doi.org/10.1063/1.522898}{\emph{J. Math. Phys.}
  {\bf 17} (1976) 303--321}.

\bibitem{Wall:2011hj}
A.~C. Wall, \emph{{A proof of the generalized second law for rapidly changing
  fields and arbitrary horizon slices}},
  \href{http://dx.doi.org/10.1103/PhysRevD.87.069904,
  10.1103/PhysRevD.85.104049}{\emph{Phys. Rev.} {\bf D85} (2012) 104049},
  [\href{https://arxiv.org/abs/1105.3445}{{\tt 1105.3445}}].

\bibitem{Bousso:2014uxa}
R.~Bousso, H.~Casini, Z.~Fisher and J.~Maldacena, \emph{{Entropy on a null
  surface for interacting quantum field theories and the Bousso bound}},
  \href{http://dx.doi.org/10.1103/PhysRevD.91.084030}{\emph{Phys. Rev.} {\bf
  D91} (2015) 084030}, [\href{https://arxiv.org/abs/1406.4545}{{\tt
  1406.4545}}].

\bibitem{Casini:2017roe}
H.~Casini, E.~Teste and G.~Torroba, \emph{{Modular Hamiltonians on the null
  plane and the Markov property of the vacuum state}},
  \href{http://dx.doi.org/10.1088/1751-8121/aa7eaa}{\emph{J. Phys.} {\bf A50}
  (2017) 364001}, [\href{https://arxiv.org/abs/1703.10656}{{\tt 1703.10656}}].

\bibitem{Lashkari:2017rcl}
N.~Lashkari, \emph{{Entanglement at a Scale and Renormalization Monotones}},
  \href{http://dx.doi.org/10.1007/JHEP01(2019)219}{\emph{JHEP} {\bf 01} (2019)
  219}, [\href{https://arxiv.org/abs/1704.05077}{{\tt 1704.05077}}].

\bibitem{Koeller:2017njr}
J.~Koeller, S.~Leichenauer, A.~Levine and A.~Shahbazi-Moghaddam, \emph{{Local
  Modular Hamiltonians from the Quantum Null Energy Condition}},
  \href{http://dx.doi.org/10.1103/PhysRevD.97.065011}{\emph{Phys. Rev.} {\bf
  D97} (2018) 065011}, [\href{https://arxiv.org/abs/1702.00412}{{\tt
  1702.00412}}].

\bibitem{Casini:2011kv}
H.~Casini, M.~Huerta and R.~C. Myers, \emph{{Towards a derivation of
  holographic entanglement entropy}},
  \href{http://dx.doi.org/10.1007/JHEP05(2011)036}{\emph{JHEP} {\bf 05} (2011)
  036}, [\href{https://arxiv.org/abs/1102.0440}{{\tt 1102.0440}}].

\bibitem{Wong:2013gua}
G.~Wong, I.~Klich, L.~A. Pando~Zayas and D.~Vaman, \emph{{Entanglement
  Temperature and Entanglement Entropy of Excited States}},
  \href{http://dx.doi.org/10.1007/JHEP12(2013)020}{\emph{JHEP} {\bf 12} (2013)
  020}, [\href{https://arxiv.org/abs/1305.3291}{{\tt 1305.3291}}].

\bibitem{Cardy:2016fqc}
J.~Cardy and E.~Tonni, \emph{{Entanglement hamiltonians in two-dimensional
  conformal field theory}},
  \href{http://dx.doi.org/10.1088/1742-5468/2016/12/123103}{\emph{J. Stat.
  Mech.} {\bf 1612} (2016) 123103},
  [\href{https://arxiv.org/abs/1608.01283}{{\tt 1608.01283}}].

\bibitem{Casini:2009sr}
H.~Casini and M.~Huerta, \emph{{Entanglement entropy in free quantum field
  theory}}, \href{http://dx.doi.org/10.1088/1751-8113/42/50/504007}{\emph{J.
  Phys.} {\bf A42} (2009) 504007}, [\href{https://arxiv.org/abs/0905.2562}{{\tt
  0905.2562}}].

\bibitem{Arias:2018tmw}
R.~E. Arias, H.~Casini, M.~Huerta and D.~Pontello, \emph{{Entropy and modular
  Hamiltonian for a free chiral scalar in two intervals}},
  \href{http://dx.doi.org/10.1103/PhysRevD.98.125008}{\emph{Phys. Rev.} {\bf
  D98} (2018) 125008}, [\href{https://arxiv.org/abs/1809.00026}{{\tt
  1809.00026}}].

\bibitem{Wong:2018svs}
G.~Wong, \emph{{Gluing together Modular flows with free fermions}},
  \href{http://dx.doi.org/10.1007/JHEP04(2019)045}{\emph{JHEP} {\bf 04} (2019)
  045}, [\href{https://arxiv.org/abs/1805.10651}{{\tt 1805.10651}}].

\bibitem{Lashkari:2015dia}
N.~Lashkari, \emph{{Modular Hamiltonian for Excited States in Conformal Field
  Theory}}, \href{http://dx.doi.org/10.1103/PhysRevLett.117.041601}{\emph{Phys.
  Rev. Lett.} {\bf 117} (2016) 041601},
  [\href{https://arxiv.org/abs/1508.03506}{{\tt 1508.03506}}].

\bibitem{Skenderis:2008dg}
K.~Skenderis and B.~C. van Rees, \emph{{Real-time gauge/gravity duality:
  Prescription, Renormalization and Examples}},
  \href{http://dx.doi.org/10.1088/1126-6708/2009/05/085}{\emph{JHEP} {\bf 05}
  (2009) 085}, [\href{https://arxiv.org/abs/0812.2909}{{\tt 0812.2909}}].

\bibitem{Botta-Cantcheff:2015sav}
M.~Botta-Cantcheff, P.~Martínez and G.~A. Silva, \emph{{On excited states in
  real-time AdS/CFT}},
  \href{http://dx.doi.org/10.1007/JHEP02(2016)171}{\emph{JHEP} {\bf 02} (2016)
  171}, [\href{https://arxiv.org/abs/1512.07850}{{\tt 1512.07850}}].

\bibitem{Marolf:2017kvq}
D.~Marolf, O.~Parrikar, C.~Rabideau, A.~Izadi~Rad and M.~Van~Raamsdonk,
  \emph{{From Euclidean Sources to Lorentzian Spacetimes in Holographic
  Conformal Field Theories}},
  \href{http://dx.doi.org/10.1007/JHEP06(2018)077}{\emph{JHEP} {\bf 06} (2018)
  077}, [\href{https://arxiv.org/abs/1709.10101}{{\tt 1709.10101}}].

\bibitem{Rosenhaus:2014woa}
V.~Rosenhaus and M.~Smolkin, \emph{{Entanglement Entropy: A Perturbative
  Calculation}}, \href{http://dx.doi.org/10.1007/JHEP12(2014)179}{\emph{JHEP}
  {\bf 12} (2014) 179}, [\href{https://arxiv.org/abs/1403.3733}{{\tt
  1403.3733}}].

\bibitem{Rosenhaus:2014zza}
V.~Rosenhaus and M.~Smolkin, \emph{{Entanglement Entropy for Relevant and
  Geometric Perturbations}},
  \href{http://dx.doi.org/10.1007/JHEP02(2015)015}{\emph{JHEP} {\bf 02} (2015)
  015}, [\href{https://arxiv.org/abs/1410.6530}{{\tt 1410.6530}}].

\bibitem{Faulkner:2014jva}
T.~Faulkner, \emph{{Bulk Emergence and the RG Flow of Entanglement Entropy}},
  \href{http://dx.doi.org/10.1007/JHEP05(2015)033}{\emph{JHEP} {\bf 05} (2015)
  033}, [\href{https://arxiv.org/abs/1412.5648}{{\tt 1412.5648}}].

\bibitem{Lewkowycz:2014jia}
A.~Lewkowycz and E.~Perlmutter, \emph{{Universality in the geometric dependence
  of Renyi entropy}},
  \href{http://dx.doi.org/10.1007/JHEP01(2015)080}{\emph{JHEP} {\bf 01} (2015)
  080}, [\href{https://arxiv.org/abs/1407.8171}{{\tt 1407.8171}}].

\bibitem{Faulkner:2015csl}
T.~Faulkner, R.~G. Leigh and O.~Parrikar, \emph{{Shape Dependence of
  Entanglement Entropy in Conformal Field Theories}},
  \href{https://arxiv.org/abs/1511.05179}{{\tt 1511.05179}}.

\bibitem{Speranza:2016jwt}
A.~J. Speranza, \emph{{Entanglement entropy of excited states in conformal
  perturbation theory and the Einstein equation}},
  \href{http://dx.doi.org/10.1007/JHEP04(2016)105}{\emph{JHEP} {\bf 04} (2016)
  105}, [\href{https://arxiv.org/abs/1602.01380}{{\tt 1602.01380}}].

\bibitem{Sarosi:2016atx}
G.~Sárosi and T.~Ugajin, \emph{{Relative entropy of excited states in
  conformal field theories of arbitrary dimensions}},
  \href{http://dx.doi.org/10.1007/JHEP02(2017)060}{\emph{JHEP} {\bf 02} (2017)
  060}, [\href{https://arxiv.org/abs/1611.02959}{{\tt 1611.02959}}].

\bibitem{Sarosi:2017rsq}
G.~Sárosi and T.~Ugajin, \emph{{Modular Hamiltonians of excited states, OPE
  blocks and emergent bulk fields}},
  \href{http://dx.doi.org/10.1007/JHEP01(2018)012}{\emph{JHEP} {\bf 01} (2018)
  012}, [\href{https://arxiv.org/abs/1705.01486}{{\tt 1705.01486}}].

\bibitem{Lashkari:2018oke}
N.~Lashkari, H.~Liu and S.~Rajagopal, \emph{{Modular Flow of Excited States}},
  \href{https://arxiv.org/abs/1811.05052}{{\tt 1811.05052}}.

\bibitem{Lashkari:2018tjh}
N.~Lashkari, H.~Liu and S.~Rajagopal, \emph{{Perturbation Theory for the
  Logarithm of a Positive Operator}},
  \href{https://arxiv.org/abs/1811.05619}{{\tt 1811.05619}}.

\bibitem{Ugajin:2018rwd}
T.~Ugajin, \emph{{Perturbative expansions of Rényi relative divergences and
  holography}},  \href{https://arxiv.org/abs/1812.01135}{{\tt 1812.01135}}.

\bibitem{Haehl:2019fjz}
F.~M. Haehl, E.~Mintun, J.~Pollack, A.~J. Speranza and M.~Van~Raamsdonk,
  \emph{{Nonlocal multi-trace sources and bulk entanglement in holographic
  conformal field theories}},
  \href{http://dx.doi.org/10.1007/JHEP06(2019)005}{\emph{JHEP} {\bf 06} (2019)
  005}, [\href{https://arxiv.org/abs/1904.01584}{{\tt 1904.01584}}].

\bibitem{Allais:2014ata}
A.~Allais and M.~Mezei, \emph{{Some results on the shape dependence of
  entanglement and R{\'e}nyi entropies}},
  \href{http://dx.doi.org/10.1103/PhysRevD.91.046002}{\emph{Phys. Rev.} {\bf
  D91} (2015) 046002}, [\href{https://arxiv.org/abs/1407.7249}{{\tt
  1407.7249}}].

\bibitem{Mezei:2014zla}
M.~Mezei, \emph{{Entanglement entropy across a deformed sphere}},
  \href{http://dx.doi.org/10.1103/PhysRevD.91.045038}{\emph{Phys. Rev.} {\bf
  D91} (2015) 045038}, [\href{https://arxiv.org/abs/1411.7011}{{\tt
  1411.7011}}].

\bibitem{Dong:2016wcf}
X.~Dong, \emph{{Shape Dependence of Holographic Rényi Entropy in Conformal
  Field Theories}},
  \href{http://dx.doi.org/10.1103/PhysRevLett.116.251602}{\emph{Phys. Rev.
  Lett.} {\bf 116} (2016) 251602},
  [\href{https://arxiv.org/abs/1602.08493}{{\tt 1602.08493}}].

\bibitem{Bianchi:2016xvf}
L.~Bianchi, S.~Chapman, X.~Dong, D.~A. Galante, M.~Meineri and R.~C. Myers,
  \emph{{Shape dependence of holographic Rényi entropy in general
  dimensions}}, \href{http://dx.doi.org/10.1007/JHEP11(2016)180}{\emph{JHEP}
  {\bf 11} (2016) 180}, [\href{https://arxiv.org/abs/1607.07418}{{\tt
  1607.07418}}].

\bibitem{Kologlu:2019mfz}
M.~Kologlu, P.~Kravchuk, D.~Simmons-Duffin and A.~Zhiboedov, \emph{{The
  light-ray OPE and conformal colliders}},
  \href{https://arxiv.org/abs/1905.01311}{{\tt 1905.01311}}.

\bibitem{Dong:2017xht}
X.~Dong and A.~Lewkowycz, \emph{{Entropy, Extremality, Euclidean Variations,
  and the Equations of Motion}},
  \href{http://dx.doi.org/10.1007/JHEP01(2018)081}{\emph{JHEP} {\bf 01} (2018)
  081}, [\href{https://arxiv.org/abs/1705.08453}{{\tt 1705.08453}}].

\bibitem{Kravchuk:2018htv}
P.~Kravchuk and D.~Simmons-Duffin, \emph{{Light-ray operators in conformal
  field theory}}, \href{http://dx.doi.org/10.1007/JHEP11(2018)102}{\emph{JHEP}
  {\bf 11} (2018) 102}, [\href{https://arxiv.org/abs/1805.00098}{{\tt
  1805.00098}}].

\bibitem{Cordova:2018ygx}
C.~Córdova and S.-H. Shao, \emph{{Light-ray Operators and the BMS Algebra}},
  \href{http://dx.doi.org/10.1103/PhysRevD.98.125015}{\emph{Phys. Rev.} {\bf
  D98} (2018) 125015}, [\href{https://arxiv.org/abs/1810.05706}{{\tt
  1810.05706}}].

\bibitem{Balakrishnan:2019gxl}
S.~Balakrishnan, V.~Chandrasekaran, T.~Faulkner, A.~Levine and
  A.~Shahbazi-Moghaddam, \emph{{Entropy Variations and Light Ray Operators from
  Replica Defects}},  \href{https://arxiv.org/abs/1906.08274}{{\tt
  1906.08274}}.

\bibitem{Belin:2019mnx}
A.~Belin, D.~M. Hofman and G.~Mathys, \emph{{Einstein gravity from ANEC
  correlators}}, \href{http://dx.doi.org/10.1007/JHEP08(2019)032}{\emph{JHEP}
  {\bf 08} (2019) 032}, [\href{https://arxiv.org/abs/1904.05892}{{\tt
  1904.05892}}].

\bibitem{deBoer:2019uem}
J.~De~Boer and L.~Lamprou, \emph{{Holographic Order from Modular Chaos}},
  \href{https://arxiv.org/abs/1912.02810}{{\tt 1912.02810}}.

\end{thebibliography}\endgroup
\end{document}